\documentclass[preprint,12pt]{elsarticle}
\usepackage{lineno,hyperref,amssymb,amsmath,amsthm}
\usepackage{footmisc}
\usepackage{comment}
\usepackage{hyperref}
\usepackage {inputenc}
\usepackage{graphicx}

\modulolinenumbers[5]

\journal{Physics Reports}

\begin{document}
\begin{frontmatter}

\begin{comment}
To include in next version. Alvaro 21/01/25 - 25/02/25 -21/04/25

Turing-Completeness and Undecidability in Coupled Nonlinear Optical Resonators. Gordon H.Y. Li, Alireza Marandi
https://arxiv.org/abs/2501.06966

Undecidability of the spectral gap in rotationally symmetric Hamiltonians. Laura Castilla-Castellano, Angelo Lucia
https://arxiv.org/abs/2410.13589v1

Undecidable problems associated with variational quantum algorithms
Georgios Korpas, Vyacheslav Kungurtsev, Jakub Mareček
https://arxiv.org/abs/2503.23723

\end{comment}

\title{Undecidability in Physics: a Review}

\author[ALC]{\'Alvaro Perales-Eceiza} 
\ead{alvaro.perales@uah.es}

\author[UCL]{Toby Cubitt} 
\ead{t.cubitt@ucl.ac.uk}

\author[NTU,CQT]{Mile Gu} 
\ead{mgu@quantumcomplexity.org}

\author[ICM,UC]{David P\'erez-Garc\'ia} 
\ead{dperezga@ucm.es}

\author[TUM,MCQST]{Michael M. Wolf}
\ead{m.wolf@tum.de}

\affiliation[ALC]{organization={Dpto. Automática, Escuela Politécnica, Universidad de Alcalá},
            %addressline={}, 
            %city={Alcalá de Henares},
            postcode={28805}, 
            state={Madrid},
            country={Spain}}
            
\affiliation[UCL]{organization={Department of Computer Science, UCL},
            %addressline={}, 
            city={London},
            %postcode={}, 
            %state={},
            country={UK}}            

\affiliation[NTU]{organization={Centre for Quantum Technologies, National University of Singapore},
            addressline={3 Science Drive 2}, 
            %city={},
            postcode={117543}, 
            %state={},
            country={Singapore}} 
            
\affiliation[CQT]{organization={Nanyang Quantum Hub, School of Physical and Mathematical Sciences, Nanyang Technological University},
            %addressline={}, 
            %city={},
            postcode={639673}, 
            %state={},
            country={Singapore}}    

\affiliation[ICM]{organization={Instituto de Ciencias Matemáticas (CSIC-UAM-UC3M-UCM)},
            %addressline={}, 
            city={Madrid},
            postcode={28049}, 
            %state={Madrid},
            country={Spain}}  
\affiliation[UC]{organization={Departamento de Análisis Matemático, Universidad Complutense de Madrid},
            %addressline={}, 
            city={Madrid},
            postcode={28040}, 
            %state={Madrid},
            country={Spain}}  
         
\affiliation[TUM]{organization={Department of Mathematics, Technical University of Munich}
            %addressline={}, 
            %city={},
            %postcode={}, 
            %state={},
            %country={}}  
            } 
\affiliation[MCQST]{organization={Munich Center for Quantum Science and Technology (MCQST)},
            %addressline={}, 
            city={München},
            %postcode={}, 
            %state={},
            country={Germany}}

\date{\today{}}

\begin{abstract}
The study of undecidability in problems arising from physics has experienced a renewed interest, mainly in connection with quantum information problems. The goal of this review is to survey this recent development. After a historical introduction,  we first explain the necessary results about undecidability in mathematics and computer science. Then we briefly review the first results about undecidability in physics which emerged mostly in the 80s and early 90s. Finally, we focus on the most recent contributions, which we divide in two main categories: many body systems and quantum information problems.

\end{abstract}
\end{frontmatter}

\tableofcontents{}

%%%%%%%%%%%%%%%%%%%%%%%%%%%%%%%%%%%%%%%%%%%%%%%%%%%%%%%%%%%%%%%%%%%%%%%%%%%%%%%%%%%%%%%%%%%%%%%%%%
%%%%%%%%%%%%%%%%%%%%%%%%%%%%%%%%%%%%%%%%%%%%%%%%%%%%%%%%%%%%%%%%%%%%%%%%%%%%%%%%%%%%%%%%%%%%%%%%%%
%%%%%%%%%%%%%%%%%%%%%%%%%%%%%%%%%%%%%%%%%%%%%%%%%%%%%%%%%%%%%%%%%%%%%%%%%%%%%%%%%%%%%%%%%%%%%%%%%%
\section{Introduction}
\label{intro}

Since Kurt G\"odel and Alan Turing introduced undecidability in mathematics and computer science in the early 20th century, there have been many attempts to extend this concept to physics.
The main idea is simple: if mathematics has (unavoidably) undecidable statements, and mathematics is the language of physics, there could be physical models that present undecidable properties.

In a note about one of those attempts, C. Bennett wrote (\cite{Bennett90}):
\begin{quotation}
Alluding to the effects on twentieth-century physics of quantum mechanics and general relativity, the cosmologist John Wheleer once remarked that the world is simpler and stranger than we can imagine. An analogous dose of strange simplicity was delivered to mathematics in the 1930s by the discoveries of G\"odel, Turing and their contemporaries in the theory of proof and computation. Yet, there have been surprisingly few applications of these ideas to physics. In particular, the phenomenon of 'undecidability' has not asserted itself in a simple, uncontrived physical setting the way that deterministic chaos, a far weaker property, has.
\end{quotation}

And some years later, R. Penrose commented (\cite{Penrose05} p378):
\begin{quotation}
 It is perhaps remarkable, in view of the close relationship between mathematics and physics, that issues of such basic importance in mathematics as transfinite set theory and computability have as yet had a very limited impact on our description of the physical world. It is my own personal opinion that we shall find that computability issues will eventually be found to have a deep relevance to future physical theory, but only very little use of these ideas has so far been made in mathematical physics\footnote{By \textit{computability issues} Penrose alluded to undecidability, as he referred to Gödel and Turing in previous paragraphs, and provided as examples of this \textit{very little use} the results of \cite{Komar64} and \cite{Geroch1986} that will be commented on in chapter \ref{UDPFirst}.}.
\end{quotation}

Sometimes simple ideas do not have simple implementations, and taking undecidability to the realm of natural phenomena described by physical laws did not turn out to be easy at all. One of the goals of this review is to show that the situation has changed since the words of Bennett and Penrose were written, as in the last decade an increasing number of central problems in physics (many of them motivated by quantum information theory), have been proven undecidable.

Quantum information theory deals precisely with the problem of how the laws of quantum physics affect the computing capabilities of a system. It is thus a very natural framework in which undecidable problems may emerge in a natural way.

But what exactly is an undecidable problem?

In this review, we take the algorithmic point of view of Turing. A problem in computer science is specified by the input (an arbitrary bit string) and a series of rules (the program) that specifies how the output depends on the input. Illustrative examples could be:
\begin{enumerate}
    \item[(a)] Input: two natural numbers (given in binary). Output: their product.
    \item[(b)] Input: an image (given by bit strings specifying the colours of its pixels). Output: Yes or No, depending on whether the image is a dog or not.
    \item[(c)] Input: the description of an algorithm and an input to it. Output: Yes or No, depending on whether the algorithm will halt or loop forever on the given input.
\end{enumerate}

If the output is just one bit (Yes/No answer), the problem is called a decision problem. Otherwise, it is called a computational problem.

The way to solve a problem is to find an algorithm that, when receiving the input, produces the corresponding output.

A decision problem is called undecidable if no algorithm solves it. Computational problems without algorithms are called uncomputable. It is important to notice here that this is not a question of efficiency. That is, we are not asking for an algorithm whose runtime is efficient (e.g.\ polynomial) with respect to the number of bits of the input. If a problem is undecidable or uncomputable, there exists \emph{no} algorithm for it, no matter how inefficient. Problem (c)~above, known as the \textit{halting problem}, was the first undecidable problem, discovered independently by A. Church and A. Turing in 1936.

Of course, to turn undecidability into a rigorous definition, we need to define first what an algorithm is. This was precisely one of the main achievements of Turing with his definition of a Turing Machine and, as it is standard by now, this is the definition we will adopt here.

There is another (related) notion of undecidability, due to G\"odel, in the context of mathematics. Roughly speaking, a mathematical statement is undecidable with respect to a set of axioms if it cannot be proven or refuted from the axioms. G\"odel showed that there are true but unprovable sentences in mathematics. Informally, \textit{provability} works ``inside'' the system, whereas \textit{truth} comes from ``outside''. To distinguish G\"odel undecidability from Turing's one, we will call such statements \textit{axiomatically independent}, instead of undecidable.

In his famous first incompleteness theorem, G\"odel proved that any reasonably powerful set of axioms will always leave some statements that are axiomatically independent.

It is not difficult to see (section \ref{UDaxiomatic} below) that, under very mild assumptions, undecidability (\`a la Turing) is in some senses a stronger notion than axiomatic independence, in the following way. If a problem is undecidable, then (infinitely) many of its instances correspond to independent statements. Therefore, even though we will comment throughout the review that particular physics problems are undecidable, it is important to notice that this in particular implies that many of their instances cannot be proven or disproven from the axioms of mathematics (whatever they are).

But what does it mean that a physical problem is undecidable?

Since mathematics describes physical systems with such unreasonable effectiveness ---as E.P. Wigner wrote in his famous essay \cite{Wigner60}--- we might think that we can directly \emph{export} to physics the undecidability that lives in mathematics. It is important to clarify that undecidability is not a feature of the physical system; it is a feature of the mathematical model we use to describe that physical system.
It might not be possible to prove that a mathematical model of a physical system has a certain property, or that the property takes a certain concrete value. Yet that property might be experimentally measurable, and when it is measured on the real physical system, some concrete value must necessarily be obtained. However, this does not imply that physical systems can solve problems that are ``beyond'' mathematical reasoning. It may simply imply that mathematical models of physical systems are often idealizations that cannot be fully attained by real physical systems. Since any finite problem is decidable simply by enumerating the solutions, an undecidable problem must necessarily conceal an infinite somewhere: infinitely many instances, infinitely many particles, infinite precision. None of these idealized limits are directly accessible experimentally. However, as we will discuss further in section~\ref{INF}, undecidability in some idealized mathematical limit of a physical system can be reflected in surprising -- sometimes even qualitatively new -- (but still in principle computable) physical properties in the real, effectively finite physical system.

The main path that has been followed to prove that a property of a physical model is undecidable resembles the way Turing proved the undecidability of the halting problem. He constructed a simulation of one machine inside another one---enabled by his invention of Universal Turing machines that can simulate any other Turing machine on arbitrary inputs. In the same vein, in order to prove the undecidability of properties of a physical system, one embeds a universal computer in it. Once embedded, a known undecidable property of the universal computer (e.g.\ the halting problem) is mapped to a property of the physical system, which therefore has to be undecidable as well.

However, natural physical systems do not resemble Turing machines very closely, so in many cases what is embedded is not directly a Universal Turing Machine, but other structures equivalent to Turing Machines with their own known undecidable properties, like Diophantine equations, tilings, or cellular automata.

In the following sections, we review these works that arguably address more and more physically relevant problems, both classical and quantum.

The review is structured as follows.

We will start (sections \ref{History} and \ref{Math}) by recalling basic facts about undecidability: its definition, history, and an account of some classic undecidable problems in mathematics and computer science that will be used later in this review.

We will then give a quick summary of the first undecidability results in physics obtained between the 60s and early 90s (section \ref{UDPFirst}), then pass to the main part of the review (section \ref{UDPRecent}) where we give a rather detailed description of the most recent undecidability results in physics. These will be divided mainly in two main categories: many body systems (section \ref{MB}) and quantum information problems (section \ref{MBQI}). We will also comment on some problems in fluid mechanics (section \ref{Fluid}) and quantum field theories (section \ref{QFT}).

We will finish this review with an outlook (section \ref{Outlook}) in which we will draw a summary diagram connecting the undecidability results in physics reviewed in sections \ref{UDPFirst} and \ref{UDPRecent} with the corresponding undecidability results in mathematics they are deduced from. We will also comment that, although undecidability results have an inherent negative flavour --they are no-go results-- they do have a positive side. We will finish with a brief discussion about the effect of noise, the notion of undecidability over the real numbers, and the tension between the infinities in mathematics and the finite physical reality.

%%%%%%%%%%%%%%%%%%%%%%%%%%%%%%%%%%%%%%%%%%%%%%%%%%%%%%%%%%%%%%%%%%%%%%%%%%%%%%
\subsection{Complementary readings}

While we are not aware of any other comprehensive review that covers in particular the plethora of undecidability results in physical systems since the 2010s, there are other excellent articles, reviews and books that we want to mention.

\textbf{K. Svozil} has explored for many years the consequences in physics and philosophy of G\"odel's and Turing's works (including as well most of the other mathematical systems discussed here) in various of his articles and books with plenty of references \cite{Svozil94, Svozil95, Svozil97, Svozil05, Svozil11, Svozil18}.

The book by \textbf{G. Chaitin, G. Doria and N. da Costa} \cite{Chaitin11} is an excellent informal introduction to undecidability in mathematics. It starts with G\"odel and Turing, focuses on algorithmic information theory, with consequences in physics (including some of the results reviewed here), and with a more speculative final chapter on open problems and future developments.

\textbf{C. Bennett} wrote a number of interesting articles explaining some results about undecidability in physics that are covered as well in this review \cite{Bennett90, Bennett95}. \textbf{J.D. Barrow} provides interesting ``informal thoughts'' about the implications of G\"odel's theorems in physics, wondering whether all their assumptions can be applied to physical theories \cite{Barrow00,Barrow11}. There are various reviews treating undecidability in physics from the point of view of algorithmic information theory, such as the articles by \textbf{C. Calude} \cite{Calude94,Calude02b,Calude05,Calude07}. One can also find in \cite{Cooper17}, edited by \textbf{Cooper and Soskova}, a recent essay by various authors about the relevance of incomputability for the real world in different disciplines including physics.

Finally, there are plenty of references about undecidability in mathematics, like the one of  \textbf{Davis} \cite{Davis77} or, more recently, the ones by \textbf{Margenstern} \cite{Margenstern00} and \textbf{Poonen} \cite{Poonen14}. They cover undecidability and universality in different mathematical systems including Turing Machines, Diophantine equations, word problems, Post systems, molecular computations, register machines, neural networks, cellular automata and tilings. \textbf{Goodman-Strauss} covers in \cite{Goodman10} most of these systems in a more informal and didactic way.

%%%%%%%%%%%%%%%%%%%%%%%%%%%%%%%%%%%%%%%%%%%%%%%%%%%%%%%%%%%%%%%%%%%%%%%%%%%%%%%%%%%%%%%%%%%%%%%%%%%%%%
%%%%%%%%%%%%%%%%%%%%%%%%%%%%%%%%%%%%%%%%%%%%%%%%%%%%%%%%%%%%%%%%%%%%%%%%%%%%%%%%%%%%%%%%%%%%%%%%%%%%%%
%%%%%%%%%%%%%%%%%%%%%%%%%%%%%%%%%%%%%%%%%%%%%%%%%%%%%%%%%%%%%%%%%%%%%%%%%%%%%%%%%%%%%%%%%%%%%%%%%%%%%%

\section{The origin. A bit of history}
\label{History}

\textbf{Euclid's \textit{Elements}} laid the foundations of mathematics in Ancient Greece and his axiomatic system remained untouched for more than two millennia. But the famous fifth postulate of geometry (\textbf{the parallel postulate}) was always under scrutiny\footnote{\label{Euclid5th}Indeed, the proof of the independence of this postulate from the remaining axioms of plane geometry can be considered the first example of a mathematical statement which is proved undecidable with respect to an accepted axiomatic system \cite{daCosta91b}.}. In the 19th century mathematicians discovered that the parallel postulate could be changed or dropped entirely, leading to non-euclidean geometries. Besides opening a new fascinating world of curved geometries, mathematicians began to think again about the foundations of mathematics, and about the possibility of establishing a definite system of axioms and rules of inference powerful enough to construct the edifice of mathematics with absolute rigor.

The revolution in the foundations of mathematics that would come in the early 20th century was initiated in the late 19th century by  \textbf{G. Cantor}, with his work on \textbf{set theory} and---to the great disturbance of the mathematical establishment of the epoch---with his infinite hierarchy of infinities. Practically ignored by his contemporaries, his goal of axiomatizing mathematics based on set theory was continued by \textbf{G. Frege}. After many years of hard work, Frege finally presented his first book in 1893. While the second one was being printed in 1903, he received a letter from B. Russell pointing to an insuperable contradiction in his system, now known as \textbf{Russell's Paradox}: \textit{The set of all sets that are not members of themselves}\footnote{Frege immediately admitted the inconsistency in his system and never recovered from it. Russell wrote: ``As I think about acts of integrity and grace, I realize that there is nothing in my knowledge to compare with Frege's dedication to truth. His entire life’s work was on the verge of completion, much of his work had been ignored to the benefit of men infinitely less capable, his second volume was about to be published, and upon finding that his fundamental assumption was in error, he responded with intellectual pleasure clearly submerging any feelings of personal disappointment. It was almost superhuman and a telling indication of that of which men are capable if their dedication is to creative work and knowledge instead of cruder efforts to dominate and be known" \cite{Irvine16}} \cite{Irvine16}.

The most prestigious mathematician at the beginning of 20th century, D. Hilbert, was a great admirer of Cantor's and Frege's work and was determined to overcome all paradoxes and difficulties. The goal of the so-called \textbf{Hilbert's program} was to axiomatize mathematics in a formal system without contradictions (\textbf{\textit{consistency}}), in which every mathematical statement could be proved true or false by following the rules of inference (\textbf{\textit{completeness}}). The task would require a tremendous effort from all mathematicians willing to contribute. B. Russell and A. Whitehead were among those taking this on, writing the monumental opus \textit{\textbf{Principia Mathematica}} with that goal.

There was a general feeling of confidence that the task would be accomplished, and in a conference in his hometown K\"onigsberg in 1930, Hilbert pronounced the famous words \textit{We must know, we will know}\footnote{The German original \textit{WIR M\"USSEN WISSEN - WIR WERDEN WISSEN} is written on his gravestone.} (in opposition to \textit{ignoramus et ignorabimus}\footnote{"We do not know and will not know" was written by Emil du Bois-Reymond, a German physiologist, in his publication "On the limits of our understanding of nature" in 1872.}).

But ironically enough, the previous day at the same conference, a young student from the University of Vienna, \textbf{Kurt G\"odel}, had presented the result that would end Hilbert's dream.

%%%%%%%%%%%%%%%%%%%%%%%%%%%%%%%%%%%%%%%%%%%%%%%%%%%%%%%%%%%%%%%%%%%%%%%%%%%%%%%%%%%%
%%%%%%%%%%%%%%%%%%%%%%%%%%%%%%%%%%%%%%%%%%%%%%%%%%%%%%%%%%%%%%%%%%%%%%%%%%%%%%%%%%%%
\subsection{G\"odel}

Using prime numbers, G\"odel managed to construct an ingenious coding system  ---the \textbf{G\"odel numbering}--- to assign a number to any symbol in a system \textit{F}, in such a way that any group of symbols ---a sentence--- is coded by a unique number. With this powerful tool, he managed to encode into a number one particular sentence \textit{G} that refers to itself: \textbf{\textit{G is not provable in the system F.}} \footnote{In the original article, G\"odel cited the analogy of his result with the liar paradox \textit{This sentence is false} \cite{Godel31}}

This was a well-formed mathematical statement encoded in a number in a formal mathematical system which includes standard arithmetic, saying about itself that it cannot be proven in that system. In this way he arrived at his \textbf{First Incompleteness Theorem}: \textit{true} sentences exist which are not provable, i.e.\ the system is not complete. (See section \ref{GodelUD} for a more formal statement of the theorem.)

John von Neumann attended G\"odel's conference in K\"onigsberg and understood immediately its tremendous impact; later worked on it; and wrote a letter to G\"odel announcing that he had obtained an important corollary. G\"odel replied that the result had been already sent for publication by himself \cite{Raatikainen25}: the \textbf{Second Incompleteness Theorem}, which proved that for any consistent system including elementary arithmetic, the consistency of the system cannot be proved in itself.

Although it took a while for the theorems to become known (and even more to be widely accepted), they are now considered amongst the greatest achievements in mathematical logic since the ancient Greeks.

%%%%%%%%%%%%%%%%%%%%%%%%%%%%%%%%%%%%%%%%%%%%%%%%%%%%%%%%%%%%%%%%%%%%%%%%%%%%%%%%%%%%
%%%%%%%%%%%%%%%%%%%%%%%%%%%%%%%%%%%%%%%%%%%%%%%%%%%%%%%%%%%%%%%%%%%%%%%%%%%%%%%%%%%%
\subsection{Church and Turing}
\label{ChurchTuring}

Meanwhile, another student in Cambridge, \textbf{Alan Turing}, was working on the \textit{Entscheidungsproblem} (decision problem), proposed also by Hilbert in 1928.

\begin{quote}
\textbf{Decision Problem:}\\
Is there an algorithm which, given a set of axioms and a mathematical statement, decides whether it is or is not provable from the axioms?
\end{quote}

In tackling this question, the first important step was to precisely define \textit{algorithm}, or the recipe to \textit{calculate}. Two people arrived at two different solutions: \textbf{Alonzo Church} with his \textbf{Lambda-calculus} \cite{Church36} and Alan Turing with his \textbf{Turing Machine} \cite{Turing36}. Both ways of computing were later proved to be equivalent by Church, Turing and S. Kleene \cite{Copeland2024}, and they served to relate the abstract mathematical concept of computation with concrete, definable processes via the Church-Turing Thesis.

\begin{quote}
	\textbf{Church-Turing Thesis}\\
	Every (human) computation can be done by a Turing machine.
\end{quote}

Even G\"odel was quite convinced of this correspondence. He wrote: ``The resulting definition of the concept of mechanical by the sharp concept of ``performable by a Turing machine" is both correct and unique. [...] Moreover it is absolutely impossible that anybody who understands the question and knows Turing's definition should decide for a different concept" \cite{Copeland19}.

Since its formulation in the 1930's there have been many attempts to prove or refute this thesis, as well as discussions of the status of this ``thesis'' and whether or not it is amenable to proof at all. No definitive resolution has been reached. It remains a thesis, not a theorem, but one widely accepted to be true by most mathematicians and computer scientists. We refer to
\cite{Svozil94,Svozil97,Nielsen04,Davis06,Chaitin11,Copeland19,Copeland2024} for extensive discussions about the Church-Turing Thesis and subsequent extensions and variants.

In the end, it turned out that, despite being equivalent, Turing's approach was more convenient and closer to the practical notion of calculating. He started by thinking of a person with pencil and paper and ended up inventing the modern computer, a universal machine able to perform any computation, and Turing Machines became the standard foundation for modern computer science theory.

Turing continued to investigate the real-world implications of his invention and considered the following \textit{ordinary} problem

\begin{quote}
	\textbf{The Halting Problem}\\
	Input: the description of an algorithm.\\  
        Output: \texttt{YES} or \texttt{NO} depending whether the algorithm halts or not.
\end{quote}
and showed that it cannot be solved by any algorithm. In other words, it is \textbf{\textit{undecidable}} whether a program will halt on a given input.

Paradoxically, in the very same article where a universal machine that can compute any algorithm was described, Turing was showing that there are things that no computer can do; that there are no algorithms for certain tasks. He formalized as well that most real numbers cannot be computed by a Turing Machine, which by the Church-Turing thesis means that they cannot be computed at all.

%%%%%%%%%%%%%%%%%%%%%%%%%%%%%%%%%%%%%%%%%%%%%%%%%%%%%%%%%%%%%%%%%%%%%%%%%%%%%%%%%%%%
%%%%%%%%%%%%%%%%%%%%%%%%%%%%%%%%%%%%%%%%%%%%%%%%%%%%%%%%%%%%%%%%%%%%%%%%%%%%%%%%%%%%
\subsection{Rice}

In 1953, \textbf{H.G. Rice} proved a strong extension of Turing halting problem for computer science \cite{Rice53}. Turing proved that in general it is impossible to know whether a program (Turing Machine) will halt or not. With a conceptually relatively simple proof, Rice showed that (informally) \textbf{in general it is impossible to know what a program does to its input} (see \cite{Hopcroft06}, \cite{Chaitin11} for a formal statement of the theorem). If you are given an arbitrary program the only way to know what it does is running it.
For instance, the questions:\\
--Does this program print ``\texttt{Hello World}"\\
--Is this computer virus harmful? \\
are undecidable.

Of course, this does not mean that these questions are not perfectly decidable for many programs; it means that there is no algorithm capable to tell what an \emph{arbitrary} program does when given the program as input. There will never be a perfect anti-virus, and debugging computer programs will never be easy.

%%%%%%%%%%%%%%%%%%%%%%%%%%%%%%%%%%%%%%%%%%%%%%%%%%%%%%%%%%%%%%%%%%%%%%%%%%%%%%%%%%%%
%%%%%%%%%%%%%%%%%%%%%%%%%%%%%%%%%%%%%%%%%%%%%%%%%%%%%%%%%%%%%%%%%%%%%%%%%%%%%%%%%%%%
\subsection{Algorithmic Information Theory}
\label{AIT}

Soon after that, a new incompleteness theorem was proven by G. Chaitin, in connection with `\textbf{Algorithmic information theory (AIT)}'; a theory founded independently by \textbf{R. Solomonoff} \cite{Solomonoff64}, \textbf{A. Kolmogorov} \cite{Kolmogorov65} and \textbf{G. Chaitin} \cite{Chaitin66} in the 1960s, that revolves around the notion of complexity.

They managed to establish a precise and formal definition of a previously vague concept using information theory and computer science\footnote{According to Chaitin, ``AIT is the result of putting Shannon's information theory and Turing's computability theory into a cocktail shaker and shaking vigorously" \cite{Calude02a}.}. Broadly speaking, the \textbf{Kolmogorov complexity}\footnote{Also known as Solomonoff-Kolmogorov-Chaitin complexity.} of a string of bits is defined as the size in bits of the smallest computer program -- \textbf{\textsl{the elegant program}}-- that produces that string as output.

\textbf{Chaitin's incompleteness theorem} states that deciding if a program is elegant is in general axiomatically independent (i.e. undecidable in the G\"odel sense). More precisely, to prove that an individual \textit{N}-bit program is elegant, you need a theory \textit{A} containing \textit{N} bits of axioms (i.e.\ having complexity \textit{N}). It turns out that any formal axiomatic system with a finite number of axioms can prove that at most finitely many programs are elegant, in spite of the fact that there are infinitely many elegant programs\footnote{While G\"odel's theorem was inspired by the liar paradox (`This statement is false'), Chaitin's result is better related to Berry's paradox: `The smallest positive integer not definable in under eleven words'. But the sentence itself contains 10 words...}.

Using this powerful idea this theory provides another precise definition of another important concept: A single string of bits is \textbf{\textit{random}} if the size in bits of the elegant program that produces it equals the size in bits of the string, i.e.\ the string cannot be compressed by any means and the most efficient way to produce it is just printing it.

By approaching incompleteness from the perspective of AIT, Chaitin showed G\"odel undecidability to be pervasive. In an informal way: with a small amount of information (your formal axiomatic system) you cannot prove things that contain more information \ldots and there are infinitely many things to prove. In Chaitin's words \cite{Chaitin07a}:

\begin{quotation} With G\"odel it looks surprising that you have incompleteness, that no finite set of axioms can contain all of mathematical truth. With Turing incompleteness seems much more natural. But with my approach, when you look at program size, I would say that it looks inevitable. Wherever you turn, you smash up against a stone wall and incompleteness hits you in the face!
\end{quotation}

Four results about undecidability in quantum physics that are based on AIT are commented on in sections \ref{MBQuantum}, \ref{Measurement} ((Brukner (2009) \cite{Brukner09}; Paterek et al. (2010) \cite{Paterek10}; Trejo et al. (2023) \cite{Trejo2023}; Purcell et al. (2024) \cite{Purcell24}). General references about AIT and Kolmogorov complexity can be found in \cite{Chaitin03}, \cite{Calude02a}, \cite{Li08}, \cite{Chaitin11}.

%%%%%%%%%%%%%%%%%%%%%%%%%%%%%%%%%%%%%%%%%%%%%%%%%%%%%%%%%%%%%%%%%%%%%%%%%%%%%%%%%%%%
%%%%%%%%%%%%%%%%%%%%%%%%%%%%%%%%%%%%%%%%%%%%%%%%%%%%%%%%%%%%%%%%%%%%%%%%%%%%%%%%%%%%
%%%%%%%%%%%%%%%%%%%%%%%%%%%%%%%%%%%%%%%%%%%%%%%%%%%%%%%%%%%%%%%%%%%%%%%%%%%%%%%%%%%%
\section{Undecidabilty in Mathematics}
\label{Math}

Since all those pioneering works, many other problems in several areas of mathematics and computer science have shown to be undecidable. We will review those which are relevant for this review in section \ref{Extensions}. But first, we need to define the notion of undecidability more rigorously.

%%%%%%%%%%%%%%%%%%%%%%%%%%%%%%%%%%%%%%%%%%%%%%%%%%%%%%%%%%%%%%%%%%%%%%%%%%%%%%%%%%%%
%%%%%%%%%%%%%%%%%%%%%%%%%%%%%%%%%%%%%%%%%%%%%%%%%%%%%%%%%%%%%%%%%%%%%%%%%%%%%%%%%%%%
\subsection{G\"odel undecidability: axiomatic independence}
\label{GodelUD}

Now we provide a more formal statement of G\"odel's theorems, explain the concept of axiomatic independence closely linked to them, and provide references for detailed explanation.

G\"odel's theorems deal with \emph{formal systems} in mathematical logic. A \emph{formal system} consists of a finite set of symbols, a grammar that defines which combinations of symbols are well-formed formulas, a set of axioms, and inference rules that describe how theorems can be proven from the axioms. Moreover, we will assume the formal system to be \emph{effectively axiomatized} in the sense that there is a mechanistic way of deciding whether a sequence of symbols is syntactically correct, represents an axiom or displays a valid proof.

A version of G\"odel's first incompleteness theorem that is entirely \emph{syntactic}, which means that it does not involve the semantic concept of `truth', is the following:

\begin{quote}
\textbf{First Incompleteness Theorem}\\
Any consistent\footnote{\emph{Consistent} means that there is no sentence such that the sentence and its negation are both provable.} formal system \textit{F} within which a certain amount of elementary arithmetic\footnote{\label{incompleteness}See \cite{Nagel01,Franzen05,Raatikainen25} to clarify what \textit{certain amount of elementary arithmetic} means in both theorems. In short, anything at least as strong as \emph{Robinson arithmetic}, a first-order logic theory with seven axioms, is sufficient. Robinson arithmetic is a fragment of the standard axiomatization of arithmetic via the Peano axioms.} can be carried out is incomplete; i.e.\ there are statements in the language of \textit{F} which can neither be proved nor disproved in \textit{F} \cite{Raatikainen25}.
\end{quote}

\begin{quote}
\textbf{Second incompleteness theorem}\\
For any consistent formal system \textit{F} within which a certain amount of elementary arithmetic\footref{incompleteness} 
can be carried out, the consistency of \textit{F} cannot be proven in \textit{F} itself \cite{Raatikainen25}.
\end{quote}

In a formal system, a statement or proposition is said to be \textbf{independent} if it cannot be proven or disproven from the axioms of the system. If we now add a semantic layer, i.e.\ an interpretation that equips sentences with meaning, this means that the truth or falsity of the statement does not follow logically from the axioms.

\textbf{Axiomatic independence} is the key concept in G\"odel’s incompleteness theorems. The first theorem states that in any interpretation of any consistent formal system that includes basic arithmetic, there are true statements that are independent.

But a new expanded formal system can be created including one or more of those statements (or their negations) as new axioms. However, that does not save the system from incompleteness, since G\"odel's first theorem also applies to this expanded system, proving that there are also independent statements in it (a processes that can be carried out \textit{ad infinitum}).

As famous historical examples of axiomatic independence, we can cite Euclid's fifth axiom (see footnote \ref{Euclid5th}), the continuum hypothesis, and the axiom of choice in set theory (see for instance \cite{cohen2008set}).

In this review we are not going to delve deeper into G\"odel's theorems and their proofs, since undecidability results in physics mostly derive from Turing's halting problem. For rigorous and didactic references on G\"odel's results see \cite{Nagel01, Franzen05, Smith07, Raatikainen25}.

%%%%%%%%%%%%%%%%%%%%%%%%%%%%%%%%%%%%%%%%%%%%%%%%%%%%%%%%%%%%%%%%%%%%%%%%%%%%%%%%%%%%
%%%%%%%%%%%%%%%%%%%%%%%%%%%%%%%%%%%%%%%%%%%%%%%%%%%%%%%%%%%%%%%%%%%%%%%%%%%%%%%%%%%%
\subsection{Turing Machines and the Halting Problem}

Throughout this review, we will see that most of the undecidability results in physics rely on an embedding of a Turing Machine into a physical system (see Fig. \ref{fig:general}).. Let us now introduce the formal definition of a Turing machine. There are several equivalent options for that. We will take the definition  from \cite{bernstein1997quantum}. For other options, we refer for instance to \cite{Sipser12}.

A (deterministic) Turing Machine (TM) is defined by a triple $(\Sigma, Q, \delta)$ where:
\begin{itemize}
    \item  $\Sigma$ is a finite alphabet with an identified blank symbol $\#$. We call $\tilde{\Sigma}= \Sigma\setminus\#$.
    \item $Q$ is a finite set of internal control states with an identified initial state $q_0$ and final (or halting) state $q_f\not = q_0$.
    \item $\delta$ is a transition function $\delta: Q\times \Sigma \rightarrow \Sigma\times Q\times \{L,R\}$.
\end{itemize}

The TM has a two-way infinite tape of cells that is used to read and write. It is indexed by $\mathbb{Z}$. The TM also has a head that moves along the tape, which reads from the tape and writes on it. At any time, only a finite number of the tape cells may contain non-blank symbols.

A \textbf{configuration} of the TM is a description of content of the tape, the location of the head and the internal state $q\in Q$.

Given a configuration, the transition function $\delta$, evaluated on the current internal state and tape symbol scanned by the head, completely specifies what to do next: which symbol from $\Sigma$ must be written on the tape in the current head position, which internal state from $Q$ the machine should transition to, and where the head should move next, either left (L) or right (R).

By convention, in the \textbf{initial configuration}:
\begin{itemize}
  \item the head is in cell $0$,
  \item the machine is in state $q_0$, and
  \item the tape has all its cells in the blank symbols except for the \textbf{input} $x\in \tilde{\Sigma}^*$, which is written on the tape in positions $0,1,2,\cdots$
\end{itemize}

The TM halts on input $x$ if it eventually enters the final state $q_f$. The number of steps a TM takes to halt on input $x$ is called the \textbf{running time}. If a TM halts, then its output is the string written on the tape from the leftmost non-blank symbol to the rightmost non-blank symbol.

It is not difficult to see that, given a TM $T$ and an input $x$, we can define a new Turing machine $T'$ which halts on a blank input if and only if $T$ halts on $x$ and, in that case, both give the same output. $T'$ is just the machine that, on a blank input, first writes $x$ on the tape and then runs as machine $T$.

It is a remarkable fact, already proven by Turing in {\cite{Turing36}} that there exist \textbf{universal Turing machines}. A universal TM is a Turing machine that can simulate the behavior of any other TM in the sense that, if given as input the description of a Turing machine $T$, it halts if and only if $T$ halts on the blank input and, in that case, its output is identical to that of $T$. This implies that a universal TM can simulate the behavior of any other TM initialized in any possible input.

We can now define the key notions for this review: \textbf{(un)decidable problems} and \textbf{(un)computable functions}.

A function $f : \tilde{\Sigma}^* \to \tilde{\Sigma}^*$ is \textbf{computable} if there exists a TM with alphabet $\Sigma$ which, on input $x\in \tilde{\Sigma}^*$, halts and leaves as output $f(x)$. A function is \textbf{uncomputable} if it is not computable.

An \textbf{(un)decidable problem} is a YES/NO decision problem on the set $\tilde{\Sigma}^*$ for some alphabet $\Sigma$, for which the decision function $f:\tilde{\Sigma}^*\rightarrow \{\text{YES, NO}\}$ is (un)computable \footnote{To make this definition match the one just given for (un)computable functions, it is enough to identify two particular alphabet symbols as YES and NO when produced as output.}. 

Note that, given a decision function, the set of all elements in $\tilde{\Sigma}^*$ whose output is YES gives a one-to-one correspondence between decision functions and subsets of $\tilde{\Sigma}^*$, usually called {\it languages}. Using this correspondence, the problem of determining the output of a decision function is the same as the problem of determining whether a given string in $\tilde{\Sigma}^*$ belongs to a language.

%%%%%%%%%%%%%%%%%%%%%%%%%%%%%%%%%%%%%%%%%%%%%%%%%%%%%%%%%%%%%%%%%%%%%%%%%%%%%%%%%%%%
\subsubsection{The Halting Problem}

The halting problem asks whether a given TM will halt on a given input. It is one of the most important problems in theoretical computer science, and its undecidability has profound implications for the limits of computation.
\begin{quote}
	\textbf{The Halting Problem}\\
	Given a Turing machine $M$ and a string $w\in \tilde{\Sigma}$, determine whether $M$ halts on input $w$.
\end{quote}

The \textbf{halting problem is undecidable}, meaning that there exists no algorithm that can solve the halting problem for all possible TM-input pairs. In other words, there is no program that can take as input a description of a TM $M$ and an input $w$, and always correctly determine whether $M$ halts on $w$.

By the comments made in the previous section, it is also undecidable to determine if a given TM $M$ will halt on the empty input, or whether a fixed universal TM will halt on a given input $w$.

It is relatively simple to provide a non-formal sketch of the \textbf{proof by contradiction}: Suppose there exist a program $H$ that can solve the halting problem for every TM $M$ with any input $w$:\\

$H(M,w)=
\begin{cases}
\texttt{YES} & \text{when }   M(w)    \text{ halts} \\
\texttt{NO} & \text{when }   M(w)    \text{ does not halt}
\end{cases}$\\

Now we can construct a new Turing machine $D(x)$ with arbitrary input $x$ that contains $H$ as a function, with the following behaviour:

\begin{itemize}
    \item If $H(x,x)$ outputs \texttt{YES}, then $D(x)$ enters an infinite loop.
    \item If $H(x,x)$ outputs \texttt{NO}, then $D(x)$ halts.
\end{itemize}

   Now, what happens if we run $D$ with its own description as input, $D(D)$?
\begin{itemize}
    \item If $D(D)$ halts, then $H$ must have output \texttt{NO} which is a contradiction.
    \item If $D(D)$ does not halt, then $H$ must have output \texttt{YES} which is also a contradiction.
\end{itemize}

Therefore, our assumption that $H$ exists must be false, and the halting problem is undecidable. For a formal proof see \cite{Hopcroft06,Sipser12}.

Note that, as in the case of G\"odel's theorems, an essential ingredient in the proof is self-reference: a TM with its own description as input, $D(D)$.

The undecidability of the halting problem is a fundamental limitation of computation, and it has implications for many areas of computer science, including software engineering, program verification, and artificial intelligence. In this review we are just considering the implications for physical systems.

%%%%%%%%%%%%%%%%%%%%%%%%%%%%%%%%%%%%%%%%%%%%%%%%%%%%%%%%%%%%%%%%%%%%%%%%%%%%%%%%%%%%
\subsubsection{Church's Theorem}
\label{Church}

In section \ref{ChurchTuring} we have mentioned that, in addition to Turing, also Church found a negative answer to the decision problem with his $\lambda$-calculus. In his 1936 paper ``An Unsolvable Problem of Elementary Number Theory''  he first provided examples in elementary number theory and topology, and later proved several undecidability theorems in problems with integer recursive functions. (All these results are usually referred together as ``Church's Theorem'' \cite{Church36}).

Although Church's result preceded Turing's, it is less frequently cited than G\"odel's and Turing's, probably because Church and Turing quickly demonstrated jointly that $\lambda$-calculus and Turing machines were equivalent, and the latter was developed as a description of a real computational machine, while the mathematical description of the former was more abstract.

However, we mention it here because Church's results were used by Shapiro \cite{Shapiro56} to prove undecidability of other important decision problems in integer recursive functions, and in turn, one of those results was used by Komar \cite{Komar64} to obtain an undecidable problem in quantum field theory (see section \ref{UDPFirst}).

%%%%%%%%%%%%%%%%%%%%%%%%%%%%%%%%%%%%%%%%%%%%%%%%%%%%%%%%%%%%%%%%%%%%%%%%%%%%%%%%%%%%
%%%%%%%%%%%%%%%%%%%%%%%%%%%%%%%%%%%%%%%%%%%%%%%%%%%%%%%%%%%%%%%%%%%%%%%%%%%%%%%%%%%%
\subsection{Relation between (Turing) undecidability and (G\"odel) axiomatic independence}
\label{UDaxiomatic}

In some sense, (Turing) undecidability implies (G\"odel) axiomatic independence. In this brief section, we will sketch this connection following the review by Poonen \cite{Poonen14}.

Consider a formal system, which is effectively axiomatized, and a decision problem with (infinitely many) instances $i$. As is usually the case, assume that for each instance of the decision problem, one can construct (algorithmically) a statement $Y_i$ so that if $Y_i$ is provable within the formal system, then the answer to instance $i$ is YES, and if its negation $\neg Y_i$ is provable, then the answer to instance $i$ is NO.

One could then design the following algorithm to solve the decision problem: Given instance $i$, recursively list all statements provable within the formal system and check whether this list contains $Y_i$ or $\neg Y_i$; the answer being YES or NO respectively. If $Y_i$ or its negation $\neg Y_i$ were provable for all $i$, this algorithm would halt and decide correctly for any instance $i$.

Therefore, if the decision problem is (Turing) undecidable, there must exist at least one statement $Y_i$ (indeed infinitely many) so that $Y_i$ is axiomatically independent of the given axiom system.

As mentioned above, we will mainly focus on (Turing) undecidability. But using this connection, one obtains (G\"odel) axiomatic independence for each one of the undecidable problems reviewed here. Moreover, if the undecidability originates from the halting problem, for any given consistent formal system one can use this connection in order to pinpoint an individual instance of the decision problem that cannot be decided on the basis of the axioms. The reason is that, by G\"odel's second incompleteness theorem, consistency of a formal system cannot be proven within that system (unless it is inconsistent). Hence, we could, at least in principle, set up an explicit Turing machine that recursively lists all theorems that are provable from the axioms and halts iff a given inconsistent statement like ``$0=1$'' appears. If the formal system is consistent,\footnote{Technically, a somewhat stronger assumption than consistency is needed here to make this argument go through; namely, that the system cannot prove any false arithmetic statements. ``Rosser's trick'' and related methods allow these types of argument to be strengthened so that they do not require any such additional assumptions, see e.g.~\cite{Oberhoff}.} then this Turing machine will never halt, but this fact will not be provable within the formal system. For  Zermelo-Fraenkel set theory with the Axiom of Choice (ZFC), the axiomatic foundation of set theory and the most commonly used foundation of mathematics, this has been worked out by Aaronson and Yedidia (2016) \cite{YedidiaAaronson} with an explicit description of a 7910-state Turing machine with one tape and a two-symbol alphabet that cannot be proved to run forever in ZFC, and improved by O’Rear (2016) \cite{ORear}.

%%%%%%%%%%%%%%%%%%%%%%%%%%%%%%%%%%%%%%%%%%%%%%%%%%%%%%%%%%%%%%%%%%%%%%%%%%%%%%%%%%%%
%%%%%%%%%%%%%%%%%%%%%%%%%%%%%%%%%%%%%%%%%%%%%%%%%%%%%%%%%%%%%%%%%%%%%%%%%%%%%%%%%%%%
\subsection{Other undecidable problems}
\label{Extensions}

In this section we will describe some other problems in mathematics and computer science beyond the halting problem that have been proved to be undecidable and that will constitute the building blocks for the undecidable problems in physics that we will survey in the forthcoming sections.

%%%%%%%%%%%%%%%%%%%%%%%%%%%%%%%%%%%%%%%%%%%%%%%%%%%%%%%%%%%%%%%%%%%%%%%%%%%%%%%%%%%%
\subsubsection{Diophantine Equations and Hilbert's Tenth Problem}
\label{Diophantine}

Diophantine equations\footnote{After the Hellenistic mathematician of the 3rd century, Diophantus of Alexandria, who studied algebraic equations.}
are polynomial equations with integer coefficients, with two or more unknowns, in which integer solutions are also sought. A simple example is the famous Pythagorean equation for triangles with a right angle
\begin{equation}
x^{2}+y^{2}=z^{2}
\end{equation}
The search for solutions and properties of different types of Diophantine equations has taken place since ancient Greece. In his famous last \textit{theorem} of 1637, Pierre Fermat stated that, for the case of three variables, there are no integer solutions of $x^n+y^n=z^n$ when the exponent is greater than two, the cases with $n=1$ (linear) or $n=2$ (Pythagorean) therefore being the only ones with integer solutions.

Making an actual theorem out of this conjecture turned out to be an arduous task, and it did not occur until 1995 when Andrew Wiles presented his celebrated proof~\cite{Wiles,Taylor}. In general, the great difficulty of proving the existence of solutions for the various types of Diophantine equations led Hilbert to include this problem in his famous 1900 list of 23 mathematical problems for the 20th century.

\begin{quote}
\textbf{Hilbert's Tenth Problem}\\
Given a Diophantine equation with any number of unknown quantities and with integral numerical coefficients: To devise a process according to which it can be determined in a finite number of operations whether the equation has an integer solution.
\end{quote}

In other words, Hilbert was asking for the existence of an algorithm to decide whether any given Diophantine equation has integer solutions.

In 1971 Matiyasevich \cite{Matiyasevich93}, using previous results by Davis, Putnam and Robinson \cite{Davis61}, proved that such an algorithm does not exist, i.e.\ \textbf{the existence of integer solutions for the Diophantine equations is undecidable}.

This important result in mathematics has been used to prove undecidable questions in different physical systems, such as integrability of classical Hamiltonians or the existence of chaos in dynamical systems (da Costa and Doria
(1991) \cite{daCosta91a, daCosta91b}); and, more recently, solvability of quantum control problems  (Bondar and Pechen (2020) \cite{Bondar2020}), and supersymmetry breaking in quantum field theory models (Tachikawa (2023) \cite{Tachikawa23})  (sections \ref{UDPFirst}, \ref{Measurement}, and \ref{QFT}, respectively).

%%%%%%%%%%%%%%%%%%%%%%%%%%%%%%%%%%%%%%%%%%%%%%%%%%%%%%%%%%%%%%%%%%%%%%%%%%%%%%%%%%%%%%%%%%%%%%%%%%
\subsubsection{Post Correspondence Problem}
\label{PCP}

The \textbf{Post Correspondence Problem} (PCP) was introduced by E. Post in 1946 \cite{Post1946} and is very often used in proofs of undecidability because it has a simpler structure than the halting problem. We follow \cite{Sipser12}, but this is also covered in \cite{Hopcroft06}, and \cite{Goodman10} for a more informal and didactic introduction.

In the PCP, we are given a finite set of \textbf{dominoes}, each with a top string and a bottom string. The goal is to determine whether there exists a sequence of dominoes such that concatenating the top strings produces the same result as concatenating the corresponding bottom strings. Formally, given an alphabet $\Sigma$ with at least two symbols, an \textbf{instance of the PCP} is a finite set of pairs $(t_i, b_i)$ --\textit{dominoes}-- over $\Sigma$: $\{ (t_1, b_1), (t_2, b_2), \ldots, (t_n, b_n) \}$,
where $t_i, b_i\in \Sigma^*$.\footnote{$\Sigma^*$ is defined as the set of all finite strings over symbols from $\Sigma$, including the empty string. }
A solution --called a \textit{match}--  is a sequence of indices $i_1, i_2, \ldots, i_k$ (with $1 \leq i_j \leq n$) such that:
$ t_{i_1} t_{i_2} \ldots t_{i_k} = b_{i_1} b_{i_2} \ldots b_{i_k}$. If such a sequence exists, we say the PCP instance has a solution; otherwise, it does not.

\begin{quote}
\textbf{Post Correspondence Problem --PCP--}\\
Given an instance of the PCP problem, decide whether its has a solution or not.
\end{quote}

By a reduction to the Turing Machine halting problem, E.~Post proved that the \textbf{PCP is undecidable}. The PCP was used as the basis to prove the undecidability of the Matrix Mortality problem which we comment below.

%%%%%%%%%%%%%%%%%%%%%%%%%%%%%%%%%%%%%%%%%%%%%%%%%%%%%%%%%%%%%%%%%%%%%%%%%%%%%%%%%%%%%%%%%%%%%%%%%%
\subsubsection{Matrix mortality}
\label{MMortality}

This problem deals with the question of whether we can determine if, for a given set of matrices $S$, there exists a finite sequence of matrices in $S$ such that their product is the zero matrix.

To make this more formal, consider a finite set of $d \times d$ matrices $S = \{M_1, \dots, M_n\} \subset M_d(\mathbb{Z})$ with integer entries. We call $S$ mortal if there is a non-empty word $w \in \{1, \dots, n\}^*$, of length $m$, say, such that for the corresponding product of matrices:
$$M_{w_1}M_{w_2} \dotsm M_{w_m} = 0.$$

\begin{quote}
\textbf{Matrix mortality problem (with parameters $(d,n)$).}
Given a set of $d \times d$ matrices $S = \{M_1, \dots, M_n\} \subset M_d(\mathbb{Z})$ with integer entries, determine whether the set is mortal.
\end{quote}

The Matrix Mortality Problem was proven to be \textbf{undecidable} even with $d=3$ and $n=16$ by Paterson in 1970 \cite{Paterson1970} with a reduction to the PCP. Improved pairs $(d,n)$ for which the Matrix mortality problem remains undecidable were obtained later by several authors \cite{blondel2000survey}.

This problem has been used to prove undecidability results in tensor networks (Kliesch et al.\ (2014) \cite{Kliesch14}, De las Cuevas et al.\ (2016) \cite{Delascuevas16}) and quantum measurement theory (Eisert (2012) \cite{Eisert12}) that are discussed in sections \ref{MBTN} and \ref{Measurement} respectively.

%%%%%%%%%%%%%%%%%%%%%%%%%%%%%%%%%%%%%%%%%%%%%%%%%%%%%%%%%%%%%%%%%%%%%%%%%%%%%%%%%%%%%%%%%%%%%%%%%%
\subsubsection{Word Problem}
\label{Word}
This problem deals with \textbf{finitely presented groups}. Given a finite alphabet $\alpha=\{a,b,c,\ldots\}$, the free group over $\alpha$ is simply the set of all different words on elements of $\alpha$ and their inverses $\{a^{-1},b^{-1}, c^{-1},\ldots\}$, with the convention that whenever an element and its inverse appear in the word in consecutive positions, the corresponding word is considered the same as the one in which both symbols are erased. For example, the word $ac^{-1}c^{-1}b^{-1}b a$ is the same as $ac^{-1}c^{-1}a$. One considers also the empty word, equivalent to the identity element $e$ in the group.
The group operation is given by word concatenation, and the inverse of a word is the inverse of its letters written in reverse order. For simplicity, one denotes $a^2=aa$, $a^3=aaa$, etc.

One can add to the picture more \textbf{relations} between the different letters, for instance $a^2b=e$. This means that, whenever one sees the block $a^2b$ in a word, one can replace it by the empty word (i.e.\ erase the block), and the new word formed by this substitution is considered the same as the initial one.

A finitely presented group is given by a finite alphabet $\alpha$ and a finite set of relations $\mathcal{S}$ on the elements of $\alpha$ and their inverses.

The word problem for a given finitely presented group can then be stated as follows:

\begin{quote}
\textbf{Word Problem.}
Given a word, decide whether it is equivalent to the empty word.
\end{quote}

In the 50s,  Novikov \cite{novikov1955algorithmic} and Boone \cite{boone1959word} independently proved that there exist finitely presented groups for which the Word Problem is undecidable.

Several of the works we will review here, like the ones of Geroch and Hartle (1986) \cite{Geroch1986}, Morton and Biamonte (2012) \cite{Morton12}, Slofstra (2019, 2020) \cite{Slofstra19,Slofstra20}, Fritz (2021) \cite{Fritz16}, Mancinska and Roberson (2020) \cite{Mancinska20} and  Fu et al.(2021) \cite{Fu21}, are based on the undecidability of the Word Problem.

%%%%%%%%%%%%%%%%%%%%%%%%%%%%%%%%%%%%%%%%%%%%%%%%%%%%%%%%%%%%%%%%%%%%%%%%%%%%%%%%%%%%%%%%%%%%%%%%%%
\subsubsection{Wang tiles}
\label{Tilings}

A \textbf{Wang tile} is a square shape with marks (typically colors) along each edge. A tiling is formed by arranging the tiles in the plane, without rotation or reflections, so that the colors on adjacent edges match. A \textbf{tiling problem} is a set of tiles $\mathcal{T}$ with which to tile the whole plane $\mathbb{Z}^2$, or a finite region of it, fulfilling those conditions.

In 1961 H.~Wang conjectured that a finite set of tiles can tile the entire plane if and only if it can tile the plane periodically (\cite{Wang61}). A corollary of this is that the decision problem of determining whether or not a given set of Wang tiles can tile the plane would be decidable. However, in 1966 his student R.~Berger disproved the conjecture by embedding a Turing Machine into a tile set in such a way that they can tile the plane if and only if the machine will never halt. Therefore \textbf{the Wang tiling problem is undecidable}. Berger's construction was substantially simplified by Robinson (1971) \cite{Robinson71}. 

With this new universal system at their disposal, researchers tried to spread undecidability by making different kinds of connections, with cellular automata being the first candidates due to their similar structure when mapping their temporal dimension to one of the spatial dimensions in the tilings (see next section). The regular structure of tilings in the plane allows as well for direct mappings to different physical systems, to prove undecidable properties in them. The results by  Kanter (1990) \cite{Kanter90} (section \ref{UDPFirst}),  Cubitt et al. (2015) \cite{Cubitt15a} and Scarpa et al.
(2020) \cite{Scarpa20} (section \ref{UDPRecent}), are based on or make use of mappings between tilings and 2-dimensional Hamiltonians.

The classical reference for tilings is the book by Gr\"unbaum and Shephard \cite{Grunbaum86}. See as well \cite{Goodman10} for an informal and didactic introduction to undecidability in tiling problems.

%%%%%%%%%%%%%%%%%%%%%%%%%%%%%%%%%%%%%%%%%%%%%%%%%%%%%%%%%%%%%%%%%%%%%%%%%%%%%%%%%%%%%%%%%%%%%%%%%
\subsubsection{Cellular Automata}
\label{CA}

Cellular Automata (CA) are simple discrete mathematical systems that evolve in discrete time steps. A CA can be characterized by its dimension, the number of states each cell can take on, the rule of evolution of each cell based on the states on neighboring cells and the range of the interaction.

They were first introduced by John Von Neumann in the 1950s as a new tool to study discrete dynamical systems particularly suitable to computer simulation. Since then, even the simpler of these apparently simple systems have displayed every kind of complex behavior, up to universal computation and undecidability.

\textbf{Conway's Game of Life} was developed by mathematician John Conway in 1970 and became known thanks to a Scientific American article \cite{Gardner70}. From a finite initial condition in the center of a 2D grid, this CA with two states (a\textit{live} and \textit{dead}) and first neighbor interactions (range $r=1$) can evolve into very diverse complex behavior, including universal computation \cite{Rendell15, Blundell16}.

But there are even simpler examples with this universal capability, such as \textbf{rule 110}: the simplest so far (1D, two-states, $r=1$). This CA was conjectured to be Turing complete in 1985 by S.~Wolfram and proved to be so in 2004 by emulating cyclic tag systems, in turn proven universal by embedding a Turing Machine into them  \cite{Cook04}.

Stephen Wolfram has been a prominent contributor to the development and study of CAs since the 1980s  \cite{Wolfram84, Wolfram85}. His book ``A New Kind of Science" \cite{Wolfram02} discusses various results about universal computation and undecidability in these and other systems, such as universality of rule 110, finding a universal Turing machine with $2$ internal states and an alphabet of size $5$, and proposing an even simpler candidate with $2$ internal states and $3$ alphabet symbols (that eventually was proved universal by the undergraduate student A.~Smith in 2007 \cite{Smith2007}).

J.~Kari \cite{Kari94} has studied undecidable properties of CA using mappings to Wang tiles to prove general results such as Rice's theorem for CAs. The application is restricted to \textit{limit sets}, which are configurations that can occur after arbitrary many computational steps and therefore represent long-term behaviour. In \cite{Kari12} Kari reviews the main results about undecidability in CAs considering diverse properties including reversibility, time-symmetry, conservation laws, chaotic dynamics, and the influence of the dimension in undecidability.

It is precisely this diversity, together with their local interactions and their regularity in space, that make CAs well-suited for embedding into planar physical systems, in which the temporal evolution of a 1D CA is mapped to a spatial dimension of the ground state of a 2D Ising lattice, for instance. We will discuss results about undecidability in physical systems obtained in this way in the next sections (Omohundro (1984) \cite{Omohundro84},  Gu et al. (2009) \cite{Gu09}, Gu and Perales (2012) \cite{Gu12}).

Let us finally point out that Israeli and Goldenfeld \cite{Israeli04,Israeli06} constructed local coarse-grained descriptions of 1D CAs that can emulate the large-scale behavior of the original systems without accounting for small-scale details. They showed that although some of the original CAs present undecidable properties, the coarse-grained large-scale construction can be predictable and even decidable at some level of description.

%%%%%%%%%%%%%%%%%%%%%%%%%%%%%%%%%%%%%%%%%%%%%%%%%%%%%%%%%%%%%%%%%%%%%%%%%%%%%%%%%%%%%%%%%%%%%%%%%
%%%%%%%%%%%%%%%%%%%%%%%%%%%%%%%%%%%%%%%%%%%%%%%%%%%%%%%%%%%%%%%%%%%%%%%%%%%%%%%%%%%%%%%%%%%%%%%%%
%%%%%%%%%%%%%%%%%%%%%%%%%%%%%%%%%%%%%%%%%%%%%%%%%%%%%%%%%%%%%%%%%%%%%%%%%%%%%%%%%%%%%%%%%%%%%%%%%
\section{Undecidability in Physics. First Results}
\label{UDPFirst}

There are undecidable questions in mathematics, the language of physics is mathematics, therefore it is possible that there are undecidable questions in physics. The conceptual similarities of this simple argument with the strange uncertainties provided by quantum mechanics at the beginning of the 20th century were soon noticed by some, but it seems that G\"odel himself did not want to explore this path\footnote{John Wheeler asked G\"odel about the connection between his incompleteness and Heisenberg’s Uncertainty Principle and ``G\"odel got angry and threw me out of his office'' \cite{Barrow11}.}.

We can consider the very Turing Machine and its halting problem as the first result of undecidability in the physical world. Except for the idealization of the infinite tape (see section \ref{INF}), it is a mechanical machine that can be perfectly built. However, Alan Turing's motivation was not to consider that connection either, but to rigorously establish the concept of an algorithm and universal machine (and as a consequence lay the foundations of theoretical computer science).

It was not until a few decades later that philosophers of science first, and mathematicians and physicists later, began to seriously explore the possibility of undecidability in physics.

Among the former, the philosopher of science \textbf{Karl Popper} \cite{Popper50} used G\"odel's results to describe a mechanical machine called `predictor' with a `calculator'\footnote{With remarkable similarities to a Turing Machine: ``A classical mechanical calculating and predicting machine which is so constructed as to produce permanent records of some kind (such as a tape with holes punched into it) capable of being interpreted as predictions of the positions, velocities, and masses of physical particles. It will be shown that such a machine can never fully predict every one of its own future states \ldots Moreover, can never fully predict, or be predicted by, any sufficiently similar machine with which it interacts.'' \cite{Popper50}} to hypothesize that ``most systems of physics, including classical physics and quantum physics, are indeterministic in perhaps an even more fundamental sense than the one usually ascribed to the indeterminism of quantum physics''. \textbf{Schlegel (1967)} \cite{Schlegel67} treated incompleteness in all sciences in the light of G\"odel's theorems (in particular uncertainty in quantum mechanics) and  \textbf{Zwick (1978)} \cite{Zwick78} and \textbf{Rothstein (1982)} \cite{Rothstein82} explored the connections with the measurement problem and irreversibility, respectively. None of them included mathematical proofs.

The first rigorous result in the field was obtained by \textbf{Komar (1964)} \cite{Komar64}, proving that the macroscopic distinguishability of two arbitrary states with infinite degrees of freedom in quantum field theory is undecidable. He mapped this system into a decision problem about distinguishability of two infinite sequences of integers that was proven undecidable via a reduction to Church's theorem \cite{Shapiro56} (section \ref{Church}).

The next batch of results were obtained two decades later, in the 1980s and 1990s. \textbf{Pour-El and Richards (1981)} \cite{Pourel81} found non-computable weak solutions for a particular three-dimensional wave equation with computable initial conditions. \textbf{Fredkin and Toffoli (1982)} \cite{Fredkin82} managed to construct a reversible ``billiard ball'' model of universal computation with hard frictionless balls that could perform all necessary logic operations by elastic collisions\footnote{\label{billiard}The authors focus on the computational capabilities of their model, so do not mention the undecidable questions inherent to any universal system of computation, as first shown by Turing. Our interest lies in the fact that they prove that a simple physical system, formally identical to the atomic model of the classical kinetic theory of perfect gases, can perform universal computation and therefore present undecidable properties.}. \textbf{Omohundro (1984)} \cite{Omohundro84} simulated an arbitrary CA with a system of partial differential equations, proving therefore that this system is able to perform universal computation as well (see footnote \ref{billiard}).

In a work aiming to find a criterion to establish whether a physical theory should be acceptable by distinguishing between \textit{computable} and \textit{measurable} quantities, \textbf{Geroch and Hartle (1986)} \cite{Geroch1986} presented as an example an observable in quantum gravity for closed cosmologies --formulated as a sum over histories--, whose expectation value is not computable\footnote{The authors recognized that the mathematics of quantum gravity had not yet been developed to the point to have a complete mathematical formulation of what the equation providing the expectation value meant.}. They proposed that the calculation should involve the question of whether two simplicial 4-manifolds are topologically identical, which was proved undecidable  via a reduction to the word problem \cite{Haken73} (section \ref{Word}). It is interesting to note that this example introduces infinite magnitudes in a \textit{natural} way since the calculation of the expected value involves the infinite sum over histories (see section \ref{INF}).

We have seen in section \ref{Tilings} that tiling problems are very natural candidates to transfer undecidability to physical systems. This was the approach that \textbf{Kanter (1990)} \cite{Kanter90} took to prove undecidable properties in a 1D spin system  (anisotropic Potts Hamiltonian) using a map to the 'domino snake' problem, which is a connectivity variation of a Wang tiling problem. The author also indicated how his arguments could be extended to higher dimensions, dynamical evolution, other spin systems, and to some spatial and temporal correlation functions in these systems. He also introduced an uncertainty principle for classical systems caused by undecidability\footnote{To our knowledge no further work was done in these directions.}.

Around the same time, \textbf{Moore (1990, 1991)} \cite{Moore90,Moore91} showed how to embed a universal Turing Machine into a continuous dynamical system of just three degrees of freedom, corresponding to the motion of a single particle in a finite-sized, three-dimensional box. His construction uses plane and parabolic mirrors in the box to modify the chaotic trajectories of the dynamical system in order to obtain the Turing Machine operations needed for universality. Virtually any question about its long-term behavior is undecidable. We refer to \cite{Bennett90,Bennett95} for a discussion of this remarkable achievement.

Taking a very different approach and mapping, \textbf{da Costa and Doria (1991)} \cite{daCosta91a, daCosta91b} (see as well \cite{Chaitin11}) encode Diophantine equations in classical mechanics and prove undecidability of questions about integrability of certain Hamiltonians and about the existence of chaos in dynamical systems. They rely on results about undecidability of the existence of roots in certain classes of real functions, obtained in turn by mapping to Diophantine equations \cite{Richardson68, Wang74}. But as the authors themselves write, ``While we have given general existential proofs concerning undecidability and incompleteness in physics, we have not considered up to now the same questions with respect to particular physically interesting systems \ldots dealing with specific situations is assuredly much more difficult than to prove general facts''. One can find an introduction to these results in \cite{Stewart91}.\textbf{ Calude et al.\ (1995)} \cite{Calude94} use the same tools to provide a couple of undecidability results about integral solutions of the one-dimensional heat equation and the electrostatic plane problem.

\textbf{da Costa et al.\ (1990)} \cite{daCosta1990} obtained undecidability results in the sense of G\"odel in general relativity. They constructed an axiomatic framework for it within Zermelo–Fraenkel set theory and showed that some statements are axiomatically independent. See also \cite{Doria2011,Chaitin11}.

\textbf{Blondel et al.\ (2001)} \cite{Blondel2001a, Blondel2001b} proved that several properties of particular classes of discrete time dynamical systems (global convergence, global asymptotic stability, mortality, and nilpotence) are undecidable by simulating Turing Machines with them.

\textbf{Lloyd (1993, 1994)} \cite{Lloyd93, Lloyd94}  pioneered fruitful interconnections between quantum information theory and undecidability, which blossomed more than 15 years later, as we will review in detail in the next section. In \cite{Lloyd93, Lloyd94} it is argued that certain spectral properties of quantum mechanical operators are undecidable. The argument is based on the following idea:  the unitary evolution $U$ associated to the time evolution of a universal quantum computer has invariant subspaces with discrete spectrum, corresponding to computations that halt, and other invariant subspaces with continuous spectrum, corresponding to computations that do not halt. Via the undecidability of the halting problem, Lloyd concludes that it is undecidable to know if the quantum state associated to a given program (in the infinite-dimensional space in which $U$ is defined) has overlap, or not, with an invariant subspace having a discrete spectrum\footnote{See \cite{lloyd2016uncomputability,cubitt2016comment} for a  discussion on the relation between \cite{Lloyd93, Lloyd94} and \cite{Cubitt15a, Cubitt15b}.}.

Due to its importance in several areas of mathematics and physics, it is interesting to note that there has been an important series of papers devoted to understanding the exact difficulty (in the sense of ``degree of uncomputability'') of several problems regarding spectra of infinite dimensional operators. One approach is classified via the so-called Solvability Complexity Index hierarchy introduced by \textbf{Hansen (2011)} \cite{hansen2011solvability} and it is crucial to understand the possibility of using some undecidable problems as part of computer assisted proofs, as has been done for instance in the solution of Kepler's conjecture \cite{hales2005proof,hales2017formal}. We will not delve into this topic here and just point to the work by \textbf{Colbrook and Hansen (2022)} \cite{colbrook2022foundations} and the references therein, also for the long history of the problem of computing spectra of operators, which can be traced back at least to \textbf{Schr\"odinger (1940)} \cite{schrodinger1940method}.

%%%%%%%%%%%%%%%%%%%%%%%%%%%%%%%%%%%%%%%%%%%%%%%%%%%%%%%%%%%%%%%%%%%%%%%%
%%%%%%%%%%%%%%%%%%%%%%%%%%%%%%%%%%%%%%%%%%%%%%%%%%%%%%%%%%%%%%%%%%%%%%%%
%%%%%%%%%%%%%%%%%%%%%%%%%%%%%%%%%%%%%%%%%%%%%%%%%%%%%%%%%%%%%%%%%%%%%%%%
\section{Undecidability in Physics. Recent Results}
\label{UDPRecent}

After those first results, there was a gap of 15 years until the recent revival of the topic, mainly driven by the development of quantum information theory, where physics and theoretical computer science are intrinsically linked.

In this (main) part of the review, we will cover undecidability results proven in the period 2008-2024.  We will group the results into two main categories: many-body systems and quantum information theory; with a couple of final paragraphs about fluid mechanics and quantum field theories.

%%%%%%%%%%%%%%%%%%%%%%%%%%%%%%%%%%%%%%%%%%%%%%%%%%%%%%%%%%%%%%%%%%%%%%%%
%%%%%%%%%%%%%%%%%%%%%%%%%%%%%%%%%%%%%%%%%%%%%%%%%%%%%%%%%%%%%%%%%%%%%%%%
\subsection{Many-Body Systems}
\label{MB}

In this section, we deal with many-body spin systems, both classical and quantum. The mathematical framework to describe those systems is the following. We consider particles located on a $D$-dimensional lattice $\mathbb{Z}^D$. To the particle in position $i$ we associate a $d$-dimensional complex Hilbert space $\mathcal{H}_i=\mathbb{C}^d$, where we denote by $|1\rangle ,\ldots |d\rangle$ its canonical (or computational) basis.
$\Lambda_L$ will denote the cube $\mathbb{Z}^D\cap [-L,L]^D$. Examples of lattices $\Lambda_L$ for $L=2$ and $D=1$ and $D=2$ are:

\
\begin{center}
\includegraphics[scale=0.3]{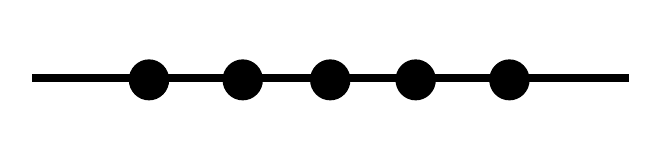} \quad \quad\includegraphics[scale=0.3]{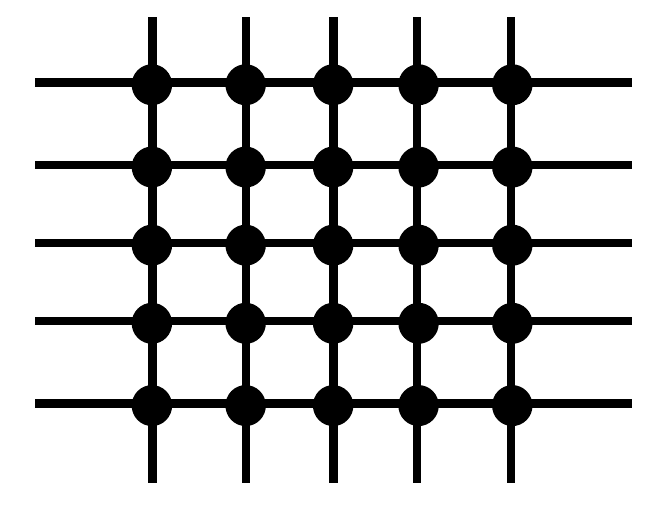}
\end{center}
\

We will restrict to the simplest case of nearest neighbor interactions between the particles, leading to a Hamiltonian of the form
\begin{equation}\label{eq:Hamiltonian-TI-boundary-terms}
H_L=\sum_{\langle i,j \rangle} h_{i,j} + \sum_{i\in \Lambda_L} k_i
\end{equation}
where $\langle i,j \rangle$ denotes neighboring positions in the lattice $\Lambda_L$, and $h_{i,j}$ (resp. $k_i$) is a Hermitian operator acting on the Hilbert space $\mathcal{H}_i\otimes \mathcal{H}_j$ (resp. $\mathcal{H}_i$).

We will also restrict to the case of translational invariant interactions, with maybe the exception of those single site terms $k_i$ for which $i$ is an element of the boundary of $\Lambda_L$. That is, in the bulk of $\Lambda_L$, all interaction terms $k_i$ will be just translations of a fixed Hermitian operator $k$ acting on $\mathbb{C}^d$, whereas for sites $i$ on the boundary, the corresponding 
$k_i$ will sometimes be allowed to vary from site to site. For the two-body terms the operators $h_{i,j}$ and $h_{i',j'}$ coincide if $i-i'=j-j'$. For instance, in 2D all horizontal interaction terms are translations of a fixed Hermitian operator $h^{\rm row}$ acting on $\mathbb{C}^d\otimes \mathbb{C}^d$ and all vertical interaction terms are translations of a fixed hermitian operator $h^{\rm col}$ acting on $\mathbb{C}^d\otimes \mathbb{C}^d$. 

The Hamiltonian is called \textit{classical} if the interactions $h_{i,j}$ and $k_i$ are diagonal in the computational basis. The eigenvalues of $H_L$ are the \textit{energy levels} of the system $\lambda_0(L)< \lambda_1(L)<\lambda_2(L)<\ldots$. In many-body spin systems, one is often interested in $\lambda_0(L)$ and $\lambda_1(L)$. Their corresponding eigenvectors are called  \textit{ground states} and \textit{elementary excitations}, respectively, and their difference $\gamma_L=\lambda_1(L)-\lambda_0(L)$ is called the \textit{spectral gap}---typically considered in the limit $L\rightarrow \infty$.

The most direct way to see that those quantities can be undecidable even for classical systems is by invoking Wang tiling problems.

As we saw in Section \ref{Tilings}, in a Wang tiling problem one is given a set of square-shaped tiles $\mathcal{T}$ with marked edges. Tiles cannot be rotated or reflected, just translated. For simplicity let us assume that the marks are colours. One is allowed to place one tile next to another one if the colours match at the common edge. With this simple rule, the tiling problem is whether one can tile, or not, the whole plane $\mathbb{Z}^2$ or, equivalently, due to Wang's extension theorem \cite{Grunbaum86}, the set of finite regions $\Lambda_L$. One can consider other versions of the tiling problem, by restricting to infinite subsets of $\mathbb{Z}^2$, like the upper-half plane. We draw here an example of a tile set with three elements and a valid tiling of size $2\times 5$:

\begin{center}
	\includegraphics[scale=0.3]{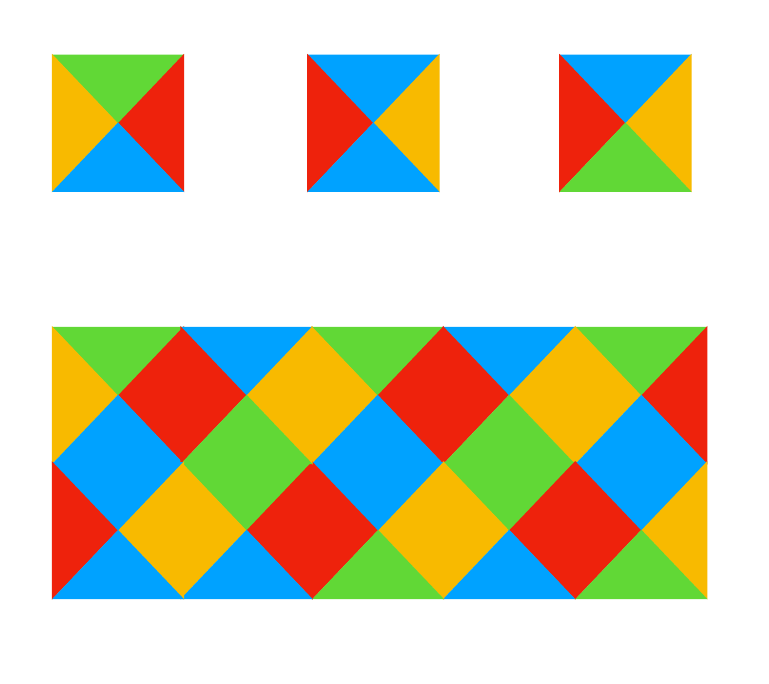}%
\end{center}\par

It is simple to convert a tiling problem into a classical many-body system as follows. The number of tiles $|\mathcal{T}|$ becomes the dimension of the local Hilbert space, and the set of tiles corresponds precisely to the set of computational basis states $|t\rangle$. On a pair of neighboring sites $i,j$, we say that a pair of tiles $(t_i, t_j)\in \mathcal{T}\times \mathcal{T}$ is not allowed if they have different colors in their common edge, that is, if they cannot be placed together in a valid tiling.  This notation allows us to define, on each pair of neighboring sites $i,j$, the projector
$$P_{i,j}=\sum_{(t_i, t_j) \text{ not allowed} } |t_i t_j\rangle\langle t_i t_j|\; ,$$
which gives rise to the (translational invariant) classical Hamiltonian
$$H_L=\sum_{\langle i,j\rangle}P_{i,j}$$
that penalizes non-allowed configurations of tiles. This construction leads to the following statement for the ground state energy $\lambda_0(L)$ of $H_L$:

{\it $\lambda_0(L)=0 $ if there exists a valid tiling of $\Lambda_L$. In that case, the set of ground states corresponds to the set of all possible valid tilings.  Otherwise $\lambda_0(L)\ge 1$.}

Since, as we saw in Section \ref{Tilings}, the tiling problem is undecidable, the same holds true for the problem of determining whether the ground state energy $\lambda_0(L)=0$ for all $L$.

Note that in this reduction from the tiling problem to the ground state energy problem, the local dimension (which is the size of the tiling set) is unbounded. However, one can easily obtain the result for a fixed local dimension  at the price of adding boundary terms which break translational invariance. For that we consider the problem of tiling the upper-half plane where one fixes the tiles in the first row. There is a straightforward way to encode the evolution of any Turing Machine in a tiling problem so that, once the first row of the tiling is fixed, there is a unique way to complete the tiling, wherein the t-th row corresponds to the configuration of the Turing Machine at time $t$ (see e.g.\ the discussion in \textbf{Robinson (1971)} \cite{Robinson71}). By removing the tiles representing transitioning to the halting state, the problem of tiling becomes equivalent to the halting problem of a Turing Machine and hence undecidable. Therefore, if we consider in this construction a Universal Turing Machine, initialize it to any input and consider the corresponding tiling to Hamiltonian reduction, it is not difficult to conclude that it is undecidable to determine whether $\lambda_0(L)=0$ for a sequence of many-body classical Hamiltonians $H_L$ as in equation (\ref{eq:Hamiltonian-TI-boundary-terms}), which are translational invariant in the bulk, have a fixed local dimension, but have non-translational invariant boundary conditions.

There are several ways in which one can improve this result that essentially dates back to the 80's \footnote{The work \cite{Kanter90} mentioned in Section \ref{UDPFirst} also has a similar flavour.} in different directions (reducing the dimension, removing the boundary terms, going to 1D, focusing on particular relevant quantities of the ground and excited states,\ldots). This is the content of the next sections.

%%%%%%%%%%%%%%%%%%%%%%%%%%%%%%%%%%%%%%%%%%%%%%%%%%%%%%%%%%%%%%%%%%%%%%%%
\subsubsection{Classical Systems}
\label{MBClassical}

For classical systems, a central question is whether one can obtain undecidability for local dimension $2$, that is for the standard setup of $\{0,1\}$-valued classical spin systems, and whether the undecidability of the ground state energy translates to relevant quantities, like the magnetization. This is precisely the content of the contributions of \textbf{Gu et al.\ (2009)} \cite{Gu09} and \textbf{Gu and Perales
(2012)} \cite{Gu12}, where the evolution of a 1D cellular automata is encoded in the ground state of a 2D (non-translationally invariant) Ising model. Using such an encoding it is shown that the {magnetization} in the ground state, its {degeneracy}, and its {correlation length} are {uncomputable quantities}.

%%%%%%%%%%%%%%%%%%%%%%%%%%%%%%%%%%%%%%%%%%%%%%%%%%%%%%%%%%%%%%%%%%%%%%%%
\subsubsection{Quantum Systems}
\label{MBQuantum}

For quantum systems, it is possible to obtain undecidability in the nearest neighbor and pure translational invariant situation and with a fixed finite local dimension. In particular, for 2D quantum many-body systems of this kind, \textbf{Cubitt et al. (2015, 2022)} \cite{Cubitt15a, Cubitt15b} showed {undecidability of the spectral gap}, the ground state energy density and other physical quantities like the existence of power-law correlations in the ground state. Moreover, they showed that undecidability of the spectral gap remains even if the considered Hamiltonians are arbitrary small perturbations of classical interactions and under the following promise: in the gapped case, the ground state is unique and a product state and the gap is of the size of the interaction strength; in the gapless case, the spectrum converges to $\mathbb{R}$ in the thermodynamic limit.

Building on the techniques developed in \cite{Cubitt15a, Cubitt15b}, this result was generalized later in several directions. It was shown to hold, with the same properties, in {1D systems} by  \textbf{Bausch et al. (2020}) \cite{Bausch20}. \textbf{Bausch et al.
(2021)} \cite{Bausch19} showed that, in 2D, {phase diagrams} are uncomputable, with both the gapped and gapless phases having non-zero measure but interleaved in an uncomputably complicated way. \textbf{ Watson
et al.\ (2022)} \cite{Watson21} showed that the {renormalization} group construction can also have an {uncomputably complex} behavior. \textbf{Cubitt (2021)} \cite{Cubitt21} showed that there is a simple, direct argument that algorithmic undecidability implies the presence or absence of a spectral gap is also axiomatically independent. \textbf{Purcell et al.\ (2024)} \cite{Purcell24} showed that even simple phase diagrams containing a single, first-order phase transition can be undecidable by constructing a system in which the phase transition occurs at an uncomputable number.\footnote{In their construction, the phase transition occurs at Chaitin's constant. This number, originally constructed by Chaitin \cite{Chaitin}, can be thought of as the probability that the universal Turing Machine will halt on a randomly chosen input, where inputs are chosen with probability inverse-exponentially proportional to their lengths.}

In a different direction, \textbf{Shiraishi and Matsumoto (2021, 2020)} \cite{Shiraishi20a,Shiraishi20b} have shown that the problem of {thermalization} in quantum many-body systems is also {undecidable}. In particular, even for the case of 1D nearest-neighbor interactions, they show that the long-time average of the expectation value of an observable (which can be taken as a spatial average of a single-site operator) is uncomputable.

%%%%%%%%%%%%%%%%%%%%%%%%%%%%%%%%%%%%%%%%%%%%%%%%%%%%%%%%%%%%%%%%%%%%%%%%
\subsubsection{Tensor Networks}
\label{MBTN}

Tensor Networks constitute a quantum information-inspired description of many-body quantum states since they aim to capture the entanglement pattern present in the state. They are one of the tools used to understand and simulate the low energy sector of quantum many-body Hamiltonians. We refer to  \cite{Cirac21} for a recent review.

Let us start with a directed graph $G=(V,E)$ and denote by $\deg^{-}(v)$ and $\deg^{+}(v)$ the ingoing and outgoing degrees of a vertex $v$ and by $\deg(v)= \deg^{-}(v)+\deg^{+}(v)$  the total degree of $v$.
A \textit{tensor network} (TN) over $G$ is defined by two ingredients. First, we associate to each edge $e$ of the graph a dimension  $D_e$ and an index $i_e$ (called a virtual index). Second, for each vertex $v$ we select a complex-valued tensor  $T^v$ with $\deg^{-}(v)$ covariant indices $i_e$, one for each incoming edge, and $\deg^{+}(v)+1$ contravariant indices $i_e$, one for each outcoming edge, plus an extra index $j_v\in \{1,\ldots, d\}$, called the physical index.
The dimension $D=\max_{e\in E}D_e$  is called the \textit{bond dimension} of the tensor network.
The tensor network $\mathcal{T}=(T^v)_{v\in V}$  defines naturally a vector $\phi_\mathcal{T}\in \bigotimes_{v\in V} \mathbb{C}^{d_v}$, which is simply the result of contracting (i.e.\ summing over, in Einstein summation convention) all edge indices $i_e$  of the tensors $T_v$.
Note that, if $N$ is the number of vertices in $G$, the number of parameters needed to describe the vector $\phi_\mathcal{T}\in \bigotimes_{v\in V} \mathbb{C}^{d_v}$ is bounded by
$$N \cdot D^{\max_{v\in V} {\rm deg}(v)}$$
which is polynomial in $N$ if the graph has bounded degree and $D={\rm poly}(N)$.
This contrasts with the exponential number $\prod_{v\in V}d_v$ required to describe an arbitrary vector $\phi \in \bigotimes_{v\in V} \mathbb{C}^{d_v}$.
Therefore tensor networks can be understood as efficient ways of describing (families of) vectors in large tensor product spaces. By a sequence of major results of Hastings \cite{hastings2006solving, hastings2007area}, improved later by several authors (see the review \cite{Cirac21}) this compressed description is still accurate enough to accurately approximate thermal states of locally interacting quantum systems, and the ground and excited states of 1d locally interacting quantum systems. This is why variational algorithms that optimize within the manifold of Tensor Network states have proven to be highly accurate in these cases.

Before going further let us give a couple of examples of tensor networks. For that, it is useful to introduce a graphical notation. Tensors will be represented by boxes and indices by legs with an orientation. Tensor contraction will then be represented just by joining an outgoing leg with an ingoing one. (In what follows we will remove the arrows from the pictures for simplicity.)

\includegraphics{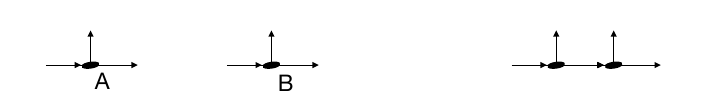}

The most basic example of a TN is a {Matrix Product State} (MPS), also known in the numerical analysis community as a {Tensor Train} \cite{Cichocki16,Cichocki17}, which has a line as its associated graph. The graphical representation of a translationally invariant MPS, where the tensor is the same in every vertex, is then

\includegraphics{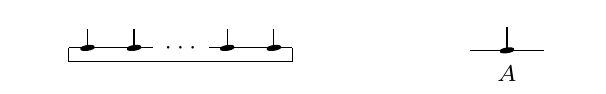}

Note that the tensor $A$ can be understood as a finite set of matrices $A^{j}$, and hence the associated Matrix Product State has the form

\begin{equation}
\label{eq:MPS-formula} \phi_A= \sum_{j_1,\ldots j_N} {\rm tr}(A_{i_1}A_{i_2} \cdots A_{i_N}) |j_1j_2 \cdots j_n\rangle
\end{equation}

Its generalization to a regular lattice is called a {Projected Entangled Pair State} (PEPS). In a square lattice, the graphical representation is depicted as follows.

\includegraphics{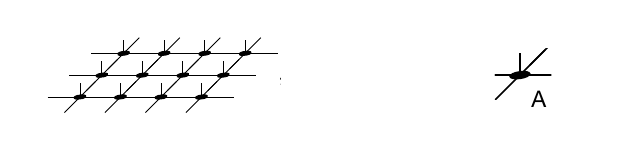}

One can re-write a tiling problem as a problem of contracting a 2D Tensor Network. For that, associated to each tiling set we define a four-leg tensor $T$ where outgoing indices $i_e$ correspond to possible colors of the corresponding side, that we denote $\ell, r, u, d$ for left, right, up, down. The value of the tensor $T_{i_\ell, i_r, i_u, i_d}$ is $1$ if the tile with colors $i_\ell, i_r, i_u, i_d$ belongs to the tile set and it is $0$ otherwise.

With this definition, it is clear that the contraction of the translationally invariant TN defined by $T$ is not equal to zero in a given region if and only if there is a valid tiling of that region, which makes the {problem of tensor contraction} of the infinite translationally invariant TN (distinguishing whether it is $0$ or $\ge 1$) undecidable.

Variants of this idea, with Word Problem as primitive instead of tiling, were reported by \textbf{Morton and Biamonte (2012)} \cite{Morton12}, where the first undecidable problems in the context of TN were analyzed.

This mapping from tiling problems to TN was explicitly used by \textbf{Scarpa et al.\ (2020)} \cite{Scarpa20}. They show that many crucial properties for the use of PEPS in quantum many-body problems, both analytically and numerically, are indeed undecidable in the thermodynamic limit. The most paradigmatic one is whether a given PEPS is {invariant under a global on-site symmetry}. That is, given a group $G$ and an associated representation $U_v(g)$ for each vertex $v$ of the graph $G$, decide whether
$\bigotimes_v U_v(g) \phi_\mathcal{T}= \phi_\mathcal{T}$ for all $g\in G$.

If one goes to the simplest one-dimensional case, one does not have access to undecidable tiling problems. Despite this, there are still important undecidable problems for the case of Matrix Product States, particularly for the case of Matrix Product Operators, where the physical index runs over the matrix units $(|j\rangle \langle k|)_{j,k=1}^d$, and therefore the vector $\phi_\mathcal{T}$ can be understood now as an operator acting on $N$ $d$-dimensional particles. Its graphical representation is simply

\begin{center}
\includegraphics[scale=0.35]{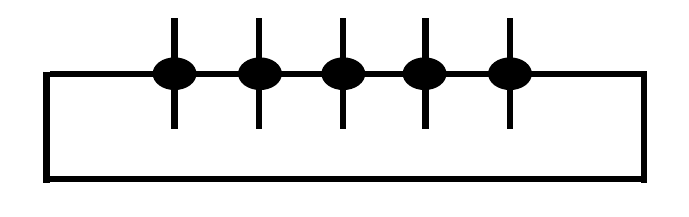}
\end{center}

In order to analyze and simulate open quantum systems with Tensor Network techniques, a central question is to know whether a given Matrix Product Operator represents a valid quantum mixed state and hence whether it is a positive semidefinite operator for all possible system sizes. In such a case it is called a Matrix Product Density Operator.

This problem has been known to be hard in practice for a long time \cite{Verstraete04} and the work-around solution was to take arbitrary Matrix Product Operators $\Theta$ and use $\Theta \Theta^\dagger$  as the variational family in the optimization algorithms. Note that, $\Theta \Theta^\dagger$ is also a Matrix Product Operator as one can easily see from the graphical notation.

Those types of Matrix Product Operators are said to have a \textit{local purification}. The problem is that they introduce an extra non-linearity in the optimization algorithm, which then becomes significantly more complex.

\textbf{Kliesch et al.\ (2014)} \cite{Kliesch14} finally proved that deciding whether a given Matrix Product Operator is positive or not for all system sizes is undecidable. Given the formula \eqref{eq:MPS-formula}, it does not come as a surprise that the result was obtained by a reduction to matrix mortality problems (see section \ref{MMortality}).

This result was used later by \textbf{De las Cuevas et al.\ (2016)} \cite{Delascuevas16} to prove the existence of Matrix Product Density Operators without a local purification. The proof uses the quantum version of Wielandt's inequality \cite{Sanz10} to show that, in case a local purification exists, there is an algorithm that finds it. Therefore, if all Matrix Product Density Operators had a local purification, checking whether a given Matrix Product Operator was positive semidefinite would be decidable, contradicting the result in \cite{Kliesch14}.

Note that {undecidability} is being used here {as proof technique} to show that an exotic object exists (in this case a Matrix Product Density Operator without a local purification). The very same idea has been used later in two remarkable directions, which we will comment on below (see section \ref{QNL}).

%%%%%%%%%%%%%%%%%%%%%%%%%%%%%%%%%%%%%%%%%%%%%%%%%%%%%%%%%%%%%%%%%%%%%%%%
\subsubsection{Open questions}

Several questions remain still open about the (un-)decidability of (quantum) many-body properties. An important one is what remains undecidable (if anything) at finite temperature, or under other sources of noise such as disorder in the interactions (see Section \ref{noise}). For instance, to the best of our knowledge the following question is still open: given a translational invariant Hamiltonian, is it decidable or undecidable to determine if the correlations in the associated thermal state decay exponentially with the distance?

%%%%%%%%%%%%%%%%%%%%%%%%%%%%%%%%%%%%%%%%%%%%%%%%%%%%%%%%%%%%%%%%%%%%%%%%
%%%%%%%%%%%%%%%%%%%%%%%%%%%%%%%%%%%%%%%%%%%%%%%%%%%%%%%%%%%%%%%%%%%%%%%%
\subsection{Quantum Information}
\label{MBQI}

It is argued by \textbf{Wolf et al.\ (2011)} \cite{Wolf11} that, due to the use of quantifiers over the integers in the first-order logic formulation of many interesting quantities in quantum information theory, it could happen that they are  uncomputable in general.  By now, this is known to be true in the following cases:

%%%%%%%%%%%%%%%%%%%%%%%%%%%%%%%%%%%%%%%%%%%%%%%%%%%%%%%%%%%%%%%%%%%%%%%%
\subsubsection{Measurements and channels}
\label{Measurement}

The link between quantum measurements and undecidability in quantum information has come in several different directions.

The first contribution in this direction, due to\textbf{ Van den Nest and Briegel (2008)} \cite{VandenNest08}, is a connection between \textit{measurement-based quantum computation} and undecidability.

In the paradigm of measurement-based quantum computation designed by  Raussendorf and Briegel \cite{Raussendorf01}, one starts with a fixed universal \textit{resource quantum state}. Then the desired computation is implemented by measuring sequentially each qubit of the resource state in a basis that depends on the quantum circuit one wants to implement.

It is known that there exist resource states for which measurement-based quantum computation is as powerful as the circuit model of computation. A central question in this area is then to understand which features of the resource state yield universal quantum computational power (see \cite{Gross07} for several results in this direction).

The family of states for which measurement-based quantum computation is understood best is the set of \textit{graph states}.

They are defined on an underlying graph where qubits initialized in $|+\rangle$ are located in the vertices of the graph and controlled-$Z$ operations are implemented on pairs of vertices joint by an edge.

If the underlying graph is a 2D square lattice, the graph state is called a \textit{cluster state}, and it is the paradigmatic universal resource for measurement-based quantum computation.

The question analyzed in \cite{VandenNest08} is the following: which property of a graph makes measurement-based computation on the associated graph state {classically simulable}. Surprisingly there is a clean answer: precisely those graphs whose associated logic, in a precise sense defined in \cite{Courcelle07}, is decidable (proved through a reduction to tiling problems \cite{Seese1991}).

A further connection between measurements and undecidability was made by \textbf{Brukner (2009)} and \textbf{Paterek et al.\ (2010)} \cite{Brukner09, Paterek10} where a link was established between randomness of outcomes of certain quantum measurements and logical independence of propositions from a given set of axioms. A somewhat related direction connecting randomness and computability culminated in the recent result by \textbf{Trejo et al.\ (2023)} \cite{Trejo2023} (see references therein for previous work). A {quantum random number generator} is proposed for which, under reasonable hypotheses, no algorithm computing any output bit can be proved correct. Experimental support is also provided.

Another connection between undecidability and quantum measurements can be traced back to the work of \textbf{Eisert et al.\ (2012)} \cite{Eisert12}, where an interpretation of matrix mortality (see section \ref{MMortality}) is given in terms of quantum measurements. The idea is the following. Any finite number of matrices $A_1, \ldots A_m$ represents a quantum measurement if we interpret them as Kraus operators of a quantum channel. That is, a measurement on state $\rho$ has outcome $j\in\{1,\ldots, m\}$ with probability
${\rm tr} (A_j \rho A_j^\dagger)$ and, in such a case, the final state becomes
$$\frac{A_j \rho A_j^\dagger}{{\rm tr} (A_j \rho A_j^\dagger)}.$$
If we repeat  the same measurement sequentially, is there a choice of outcomes that will never occur? This is the \textit{quantum measurement occurrence problem} defined in \cite{Eisert12}. Note that this is equivalent to asking whether there exist $N\in \mathbb{N}$ and $j_1,\ldots j_N\in \{1,\ldots, m\}$ so that $$A_{i_1}A_{i_2}\cdots A_{i_N}=0$$ which is precisely the matrix mortality problem and hence undecidable.

As observed by \textbf{Wolf et al.\ (2011)} \cite{Wolf11}, using the work of \textbf{Hirvensalo (2007)} \cite{Hirvensalo07}, matrix mortality problems are also undecidable if one restricts to quantum channels\footnote{Indeed classical channels suffices.}. The associated result is the \textit{reachability problem}, which is a problem in \textit{quantum control theory}.
Given an initial state $\rho$, a target state $|\psi\rangle$, a parameter $0<\lambda<1$, and a set of quantum controls $T_1,\cdots T_n$ (that are quantum channels), it is undecidable to know whether there is a choice of controls so that
\begin{equation}\label{eq-reachability}
\langle \psi| T_{i_N} \cdot T_{i_1}(\rho)|\psi\rangle>\lambda\end{equation}

This connection between quantum control and undecidability was explored further by \textbf{Bondar and Pechen (2020)} \cite{Bondar2020}.

 \textbf{Gimbert and Oualhadj (2010)} \cite{Gimbert10} gave a result about the undecidability of control problems that is much stronger than \eqref{eq-reachability}.
By a gap amplification technique, they show that, even  with the promise that \begin{itemize}
    \item either  $\langle \psi| T_{i_N} \cdots T_{i_1}(\rho)|\psi\rangle<\frac{1}{2}$ for all $N$ and all $i_1,\cdots i_N$;
    \item or for each $\epsilon>0$ there is an $N$ and $i_1,\ldots i_N$ so that $\langle \psi| T_{i_N} \cdots T_{i_1}(\rho)|\psi\rangle>1-\epsilon$,
\end{itemize}
it is undecidable to know which is the case

This result was used by \textbf{Elkouss and P\'erez-Garc\'ia (2018)} \cite{Elkouss16} to show that given a {classical channel with memory} (the result trivially holds then for the quantum case) and the promise that its {capacity} is $1$ or $<\frac{1}{2}$, it is undecidable to know which is the case. This was generalized later by  \textbf{Boche et al. (2020)} \cite{boche2020shannon}, with completely different techniques, to cover the minimal possible channel size: one bit of input, one bit of output, one bit of memory.

Let us recall that a channel with memory is just a channel with two inputs (the memory and the actual input) and the two associated outputs. It can therefore be represented by the following tensor

\begin{center}
\includegraphics[scale=0.25]{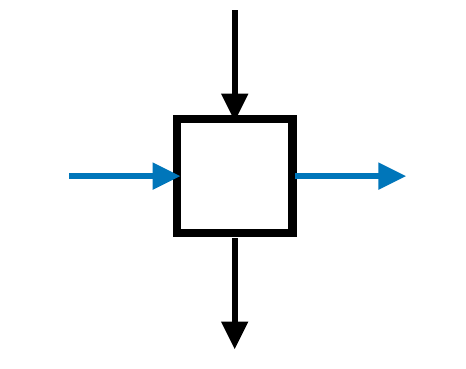}
\end{center}
where the black arrows represent the input-output of the channel and the blue ones the input-output of the memory.

Using the channel $N$ times is therefore represented by the corresponding Matrix Product Operator

\begin{center}
\includegraphics[scale=0.25]{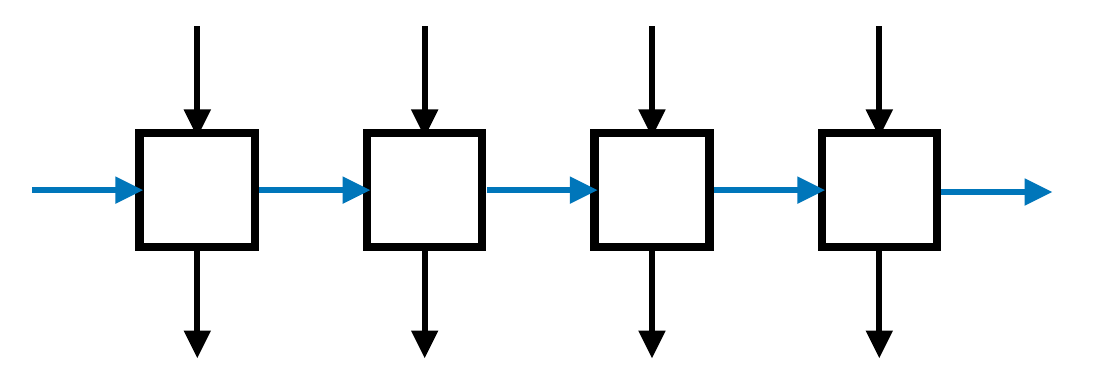}
\end{center}

The capacity of the channel is defined similarly as in the memoryless case: the optimal rate (number of bits per use of the channel) at which information can be sent if it is optimally encoded before sending.

There have been two more results regarding {undecidability in classical information theory} that we want to mention:

\textbf{Boche and M\"onich (2021)} \cite{boche2021algorithmic} proved that the {bandwidth of a bandlimited signal is uncomputable}. The relevance of this result stems from the fact that, according to the Shannon sampling theorem, the bandwidth determines the minimum sampling rate which is required for a perfect reconstruction.

\textbf{Boche and Deppe (2021)} \cite{boche2021computability} proved that the {zero-error capacity of a memoryless channel is uncomputable}. To define the zero-error capacity, there is no probability of error allowed in the transmission of information, as opposed to the standard setup where an error is tolerated, as long as it vanishes asymptotically (in the number channel uses).

However, this result comes with a caveat that is worth mentioning as it is not uncommon to results about ``uncomputable'' functions on spaces over the field of computable real numbers. It is a  jarring fact about the field of computable numbers that it is not computably ordered \cite{Rice54}. That is, statements like $x=0$ or $x>1$ are undecidable for computable real numbers in general. As a consequence, every discontinuous function, even a simple step function, becomes uncomputable.
 In fact, this is precisely the case for the zero-error capacity---it is discontinuous by the nature of its definition. 

In such a case, one could argue that the uncomputability might be more a statement about how the input is presented than a signature of any quality of the considered property that goes beyond discontinuity.

To circumvent this issue, most of the results mentioned in this review work with computably ordered subfields like the fields of rational or algebraic numbers. In the case of the zero-error capacity, one could modify the question and, for instance, consider the decision problem of whether the capacity is larger than a given value in cases where everything takes place within a computably ordered field (e.g. the algebraic numbers). To the best of our knowledge, it is currently open whether this problem is (un-)decidable.

%%%%%%%%%%%%%%%%%%%%%%%%%%%%%%%%%%%%%%%%%%%%%%%%%%%%%%%%%%%%%%%%%%%%%%%%
\subsubsection{Quantum non-locality}
\label{QNL}
We finish this section with some remarkable undecidability results in quantum information theory that deal with quantum non-locality or, equivalently, Bell inequalities.

{Bell inequalities} date back to the famous work of \cite{einstein1935can}, where Einstein, Podolsky and Rosen raise an apparent dilemma about the completeness of quantum mechanics as a model for nature. 30 years later, Bell \cite{bell1964einstein} proposed an idealized experiment to test that quantum mechanics is in fact not compatible with a classical explanation of Nature in terms of local hidden variables. Implementing this experiment was a challenging task since one needed to avoid some crucial mismatches --called {\it loopholes}--  between the required ideal setup and the actual experiment. The two main loopholes to deal with were the locality loophole and the detection loophole \footnote{There are many other loopholes one could consider, as one increases the level of scrutiny regarding potential ways nature might conspire to invalidate the Bell experiment.  As A. \ Aspect says \cite{aspect2015closing}: ``Yet no experiment, as ideal as it is, can be said to be totally loophole-free [\ldots]. Since all events have a common past if we go back far enough in time --possibly to the Big Bang-- any observed correlation could be justified by invoking such an explanation.''.}. The locality loophole refers to the need to avoid causality relations between the spatially separated measurement setups of the experiment. Solving the locality loophole requires fast random measurement choices in order to avoid any causal influence of the measurement choice at one side on the outcome at the other side. The detection loophole refers to the problem that only a small fraction of the photons are actually detected and that this fraction could depend on the configuration of the  measurement setups. 

A series of experiments initiated by J. Clauser, A. Aspect and co-authors in the 70's culminated in the 1982 experiment led by A.\ Aspect \cite{Aspect82a,Aspect82b} which removed the locality loophole, and is widely viewed as the first to convincingly demonstrating Bell inequality violation and separate quantum physics from any local hidden variable theory. Finally, in 2015 an experiment removing simultaneously the locality and detection loopholes was performed in Delft~{\cite{Hensen15}}, and independently also in Vienna~\cite{Vienna} and NIST~\cite{Boulder}, again obtaining the predicted violations of Bell inequalities. 

At the practical level, violations of Bell inequalities are behind some important applications of quantum information, such as quantum cryptography \cite{zapatero2023advances} and the secure generation of random numbers \cite{pironio2010random,liu2018device}.

A common way to present and study Bell inequalities in quantum information theory is via \textit{non-local games}. In a non-local game, a referee $R$ samples a pair of questions $x,y,$ with a known probability distribution $\pi(x,y)$ from a finite set of $N$ possible questions. He sends question $x$ to Alice and question $y$ to Bob, the two players of the (cooperative) game. They can pre-agree on a common strategy, but once the game starts, Alice and Bob cannot communicate with each other. Alice (resp. Bob) must provide an answer $a$ (resp. $b$) to the referee from a finite set of $M$ possible answers. The referee checks according to a known verification function $V(a,b,x,y)\in \{0,1\}$ whether the players win ($V(a,b,x,y)=1$) or lose the game.

\begin{center}
\includegraphics[scale=0.25]{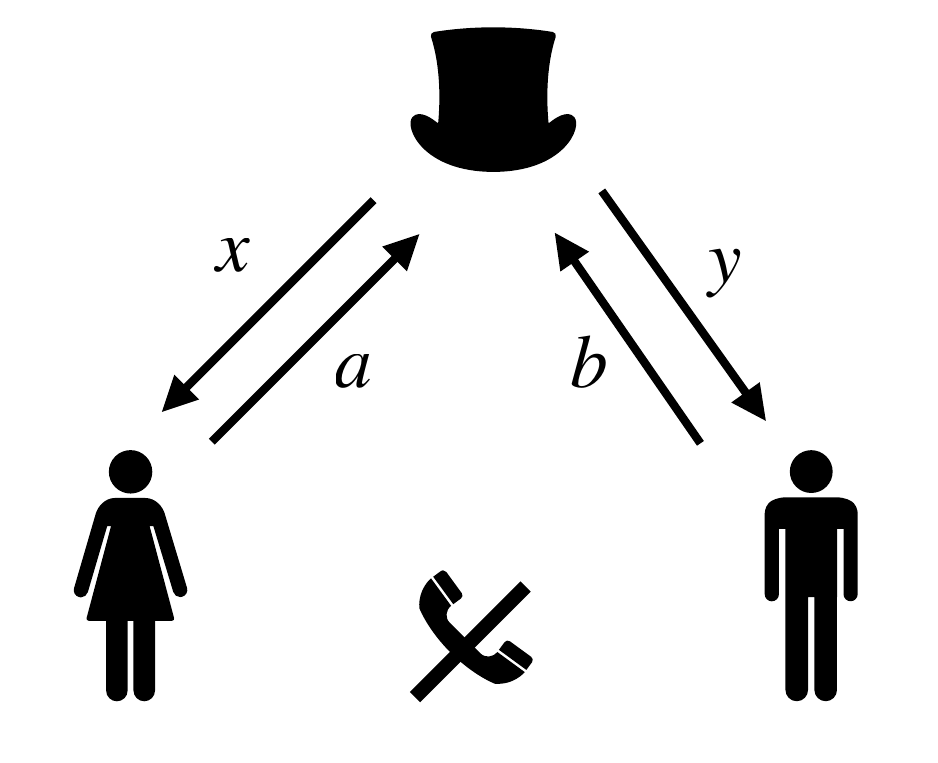}
\end{center}

The game $G$ is then given by the pair $(\pi, V)$ and its size is $\max\{N,M\}$. The value of the game is the optimal probability of winning, optimized among all allowed strategies\footnote{There are different possible variants. For instance, one can also take $V(a,b,x,y)$ to be real-valued (interpret it e.g.\ as the amount of money the players win if they answer $(a,b)$ on input $(x,y)$) and the value of the game would then be the maximum expected gain, when optimized over all possible strategies; one can add more players, or more rounds of communication between referee and players; one can make questions and/or answers quantum;  one can allow for some type of communication between the players once the game starts; \ldots. We refer to \cite{buhrman2010nonlocality,palazuelos2016survey} for more details on the topic.}.

A \textit{strategy} is simply a probability distribution $p(ab|xy)$ giving the probability of answering $(a,b)$ on question $(x,y)$. Which strategies are allowed depends on the resources given to the players. If they play using only classical physics, $p(ab|xy)$ will take the form of a hidden-variable model:
\begin{equation}\label{HVM-strategy}
p(ab|xy)= \sum_\lambda p_\lambda p^A(a|x\lambda) p^B(b|y\lambda)
\end{equation}
That is, before the game starts they sample a random variable $\lambda$ according to probability $p_\lambda$, then each one of them answers according to a probability distribution that depends only on their question and the value of $\lambda$.

The \textit{classical value} of the game $\omega (G)$ is the optimal probability of winning if the players use classical strategies. That is
$$\omega(G)= \sup_{p(ab|xy)\text{ as in \eqref{HVM-strategy}} } \sum_{xyab} \pi(x,y) V(a,b,x,y) p(ab|xy)$$

If players are allowed to share an entangled state before the game starts and can decide their answers based on some measurement made on this state, the associated value is called the \textit{entangled value} of the game $\omega^*(G)$.

There is however an important subtlety here regarding how the spatial separation between Alice and Bob is modeled. In principle, this gives rise to two different types of strategies.

If we model spatial separation using a tensor product structure, as is usual in quantum information theory, the allowed strategies are represented by

\begin{equation}\label{Quantum-strategy}
p(ab|xy)={\rm tr} \left[(E^x_a\otimes F^y_b)\rho_{AB}\right],
\end{equation}
where $\mathcal{H}_A$ and $\mathcal{H}_B$ are finite dimensional Hilbert spaces for Alice and Bob, respectively, $\rho_{AB}$ is a mixed state on $\mathcal{H}_A\otimes \mathcal{H}_B$, and $E^x_a$ (resp.\ $F^y_b$) is a generalized measurement for Alice (resp. Bob). That is, $E^x_a$ are  positive semidefinite operators acting on $\mathcal{H}_A$ so that
$\sum_a E^x_a= {\rm Id}$ for all $x$. The entangled value $\omega^*(G)$ refers to those types of strategies and is defined as follows:
\begin{equation}\label{eq:entangled-value}
\omega^*(G)= \sup_{p(ab|xy)\text{ as in \eqref{Quantum-strategy}} } \sum_{xyab} \pi(x,y) V(a,b,x,y) p(ab|xy)
\end{equation}

However,  spatial separation can also be modeled in the spirit of algebraic quantum field theory. In this case, there is a single {\it infinite} dimensional Hilbert space $\mathcal{H}$ on which both Alice's and Bob's operators act, and spatial separation is reflected by imposing that Alice's and Bob's operators commute. This is the commuting-operator paradigm that gives as possible strategies
\begin{equation}\label{Commuting-strategy}
p(ab|xy)={\rm tr} \left[(E^x_a F^y_b)\rho\right],
\end{equation}
where $\rho$ is a state on an infinite dimensional Hilbert space $\mathcal{H}$,  and $E^x_a$ and $F^y_b$ are generalized measurements on $\mathcal{H}$ so that $[E_a^x, F^y_b]=0$ for all $x,y,a,b$. We can then define the \textit{commuting entangled value}:

$$\omega_c^*(G)= \sup_{p(ab|xy)\text{ as in \eqref{Commuting-strategy}} } \sum_{xyab} \pi(x,y) V(a,b,x,y) p(ab|xy)$$

Note that, by definition $\omega(G)\le \omega^*(G)\le \omega^*_c(G)$ since the former strategies are special cases of the latter.

Bell inequalities deal with the first of those inequalities $\omega(G)\le \omega^*(G)$. A Bell inequality violation in the language of non-local games is nothing but a game $G$ for which the inequality is strict $\omega(G)< \omega^*(G)$, meaning that quantum strategies are strictly better than classical ones for that particular game.

Whether  $\omega^*(G)= \omega^*_c(G)$ for all games or not has been an important open question, known as {Tsirelson's problem}. It asks whether or not both ways to model spatial separation give the same experimental predictions.

In a breakthrough paper  \textbf{Slofstra (2020)} \cite{Slofstra20} showed that it is {undecidable to know if $\omega^*_c(G)=1$} for a given game $G$. In his subsequent paper \textbf{Slofstra (2019)} \cite{Slofstra19} showed the {same result for $\omega^*(G)$}. The proof goes via a clever mapping between finitely presented groups and a class of non-local games. The results are then obtained by invoking the undecidability of the word problem for groups (see section \ref{Word}). Slofstra's results were used later by \textbf{Fritz (2021)} \cite{Fritz16} to show that {quantum logic is undecidable}. They were also used by \textbf{Man{\v{c}}inska and Roberson (2020)} \cite{Mancinska20} to give a striking undecidability result in the context of {graph theory}. They show that, given two graphs $G, H$, knowing whether there exists a planar graph $K$ for which ${\rm hom} (K,G)\not = {\rm hom} (K,H)$ is an undecidable problem. Here ${\rm hom} (K,G)$ is the number of graph homomorphisms from $K$ to $G$. The reason underlying this undecidability is indeed the main result of \cite{Mancinska20}: ${\rm hom} (K,G)= {\rm hom} (K,H)$ for all planar graphs $K$ if and only if $G$ and $H$ are {quantum isomorphic} (see {\cite{Atserias19}} for the formal definition). This in turn can be characterized by the existence of a perfect commuting strategy for a particular non-local game defined in terms of $G$ and $H$. The result then can be obtained as a reduction from \cite{Slofstra20}. \textbf{Fu et al. (2021)} \cite{Fu21} strengthened the results of \cite{Slofstra19, Slofstra20} by showing that undecidability is kept even for a particular fixed game size.

Finally, in a breakthrough paper, \textbf{Ji et al.\ (2020)} \cite{Ji20} managed to prove that, even {under the promise that $\omega^*(G)=1$ or $\omega^*(G)\le 1/2$, it is undecidable to know which is the case}. This has the remarkable consequence that there exist games for which $\omega^*(G)\not = \omega_c^*(G)$, giving a {negative answer to Tsirelson's problem}.
The way to obtain this consequence is by using undecidability as a proof ingredient, as already commented in section \ref{MBTN}. One considers an algorithm that interleaves upper bounds to $\omega^*_c(G)$ by using the semidefinite-program-hierarchy of \textbf{Navascu\'es et al.\ (2008)} \cite{Navascues08}, with lower bounds to $\omega^*(G)$ by brute-force computing the supremum in \eqref{eq:entangled-value} for larger and larger Hilbert space dimensions. If  $\omega^*(G)= \omega^*_c(G)$, on input a game for which  $\omega^*(G)=1$ or $\omega^*(G)\le 1/2$, such an algorithm would always halt and give the right answer, making the problem decidable, which is the desired contradiction. We refer to section 8 in \cite{Fitzsimons19} for a detailed explanation.

Using a connection made in \cite{Junge11} and \cite{Fritz12} between Tsirelson's problem and Connes' embedding problem, the result in \cite{Ji20} implies also a {negative solution to Connes' embedding problem}. The latter was a major open problem in the theories of von Nuemann algebras and $C^*$-algebras that is finally solved using this completely unexpected approach based on undecidability.

Another consequence of Ji et al.\ (2020) \cite{Ji20} is the undecidability of  the membership problem for the set of quantum contextual behaviors. This was obtained in \textbf{Wright and Farkas (2023)} \cite{wright2023invertible} via the construction of an invertible map between non-locality and contextuality scenarios.

In a slightly different direction, \textbf{Bendersky et al.\ (2017)} \cite{Bendersky16} analyzed the maximum set of strategies $p(ab|xy)$ that is compatible with special relativity, in the sense that the output $a$ (resp. $b$) cannot contain information about the input $y$ (resp. $x$). That is, $\sum_{b} p(ab|xy)$ is independent of $y$ and $\sum_{a} p(ab|xy)$ is independent of $x$. The set of such strategies is called non-signaling and it contains all quantum ones.

By using undecidability results about learnability of functions (in turn reduced to the halting problem in \cite{Zeugmann08}),  it is proven in \cite{Bendersky16}  that non-signaling deterministic models for strategies $p(ab|xy)$ have to be uncomputable.

%%%%%%%%%%%%%%%%%%%%%%%%%%%%%%%%%%%%%%%%%%%%%%%%%%%%%%%%%%%%%%%%%%%%%%%%
\subsubsection{Open problems}

There are several important questions that remain open. Arguably the most important one is whether the classical or quantum capacities of a memoryless quantum channel are uncomputable. If so, this would imply that there cannot exist any {\it reasonable} single-letter formula to compute them, in analogy to Shannon's coding theorem for the capacity of a classical channel. (Note that deciding if the classical capacity of a quantum channel is non-zero -- i.e.\ whether it has any capacity at all -- is straightforward, though computing its value is not. For the quantum capacity, even deciding whether a channel has any quantum capacity at all may be non-trivial.)

Currently, the known coding theorem for the quantum capacity in terms of the coherent information $I_{\rm coh}$ requires a regularization. That is, given a quantum channel $T$, its quantum capacity is given by
$$\mathcal{Q}(T)=\lim_n \frac{1}{n} \max_{\rho_n} I_{\rm coh}(T^{\otimes n}, \rho_n).$$ Similarly, the classical capacity of a quantum channel is given by regularizing the Holevo quantity $\chi$
$$\mathcal{C}(T)=\lim_n \frac{1}{n} \max_{\rho_n} \chi(T^{\otimes n}, \rho_n).$$

\textbf{In Cubitt et al.\ (2015)} \cite{Cubitt15c}, some evidence is given concerning the uncomputability of the quantum capacity. It is shown that, for any given $n$, there exist channels $T$ with positive quantum capacity that, however, fulfill

$$\frac{1}{n} \max_{\rho_n} I_{\rm coh}(T^{\otimes n}, \rho_n)=0.$$

%%%%%%%%%%%%%%%%%%%%%%%%%%%%%%%%%%%%%%%%%%%%%%%%%%%%%%%%%%%%%%%%%%%%%%%%
%%%%%%%%%%%%%%%%%%%%%%%%%%%%%%%%%%%%%%%%%%%%%%%%%%%%%%%%%%%%%%%%%%%%%%%%
\subsection{Fluid mechanics}
\label{Fluid}
\textbf{Tao (2016)} \cite{Tao16} started a program  to address the regularity problem for Navier-Stokes equations in fluid mechanics (one of the Clay's millennium problems) by showing that fluid mechanics is Turing complete \cite{Tao17,Tao18,Tao19,Tao19b}. This would allow encoding a self-replicating machine that blows up in finite time, into the evolution of a fluid. Indeed, Tao showed this to be possible for an artificial modification (an averaged version) of the Navier-Stokes equations \cite{Tao16}.

A major achievement of that program appeared in a work by  \textbf{Cardona et al.\ (2021)} \cite{Cardona20}, where they managed to show that there exist Turing complete fluid flows on a Riemannian 3-dimensional sphere. This is done however in the simpler case of Euler equations, which model {\it inviscid} incompressible fluids, as opposed to the viscid case treated by the Navier-Stokes equations.

The particular result proven in \cite{Cardona20} is the existence of fluid flows $X$ in the $3$-dimensional sphere $S_3$ so that the following problem is equivalent to the halting problem of a Turing Machine:  Given a point $p\in S_3$ and an open set $U\subset S_3$, decide if the orbit of $X$ through $p$ intersects $U$.

The proof uses deep results of contact topology, together with the results of Moore \cite{Moore90, Moore91} commented in Chapter \ref{UDPFirst}, in his construction of uncomputable dynamical systems.

\textbf{Cardona et al.\ (2023)} \cite{cardona2023computability} have managed to extend the validity of their result to the Euclidean space $\mathbb{R}^3$.

%%%%%%%%%%%%%%%%%%%%%%%%%%%%%%%%%%%%%%%%%%%%%%%%%%%%%%%%%%%%%%%%%%%%%%%%
%%%%%%%%%%%%%%%%%%%%%%%%%%%%%%%%%%%%%%%%%%%%%%%%%%%%%%%%%%%%%%%%%%%%%%%%
\subsection{Quantum Field Theory}
\label{QFT}

We already commented in Chapter \ref{UDPFirst} on the pioneering work of Komar \cite{Komar64} about undecidability in quantum field theories. Using a different approach \textbf{Tachikawa (2023)} \cite{Tachikawa23} proved that it is undecidable whether a given 2D supersymmetric Lagrangian theory breaks supersymmetry or not by making a connection with the undecidability of Diophantine equations (section \ref{Diophantine}).

%%%%%%%%%%%%%%%%%%%%%%%%%%%%%%%%%%%%%%%%%%%%%%%%%%%%%%%%%%%%%%%%%%%%%%%%
%%%%%%%%%%%%%%%%%%%%%%%%%%%%%%%%%%%%%%%%%%%%%%%%%%%%%%%%%%%%%%%%%%%%%%%%
%%%%%%%%%%%%%%%%%%%%%%%%%%%%%%%%%%%%%%%%%%%%%%%%%%%%%%%%%%%%%%%%%%%%%%%%
\section{Outlook}
\label{Outlook}

We have first summarized in this review the concept of undecidability in mathematics from the pioneering works of G\"odel and Turing to the extensions to other mathematical systems that later served to construct mappings to physical models. In Chapter \ref{UDPFirst} we have briefly reviewed the first results of undecidability in physics. In most cases, they were models built ad hoc to embed in them a Universal Turing Machine or other equivalent systems, but that, in general, bear little resemblance to physical systems present in nature. The core of this review was Chapter \ref{UDPRecent}, where we presented the results of the last decade in which undecidable properties have been proven in more usual models in physics, both classical and quantum, but with special relevance in the latter due to the developments in quantum information theory.

As in any review, there are several topics that we left out. There is still a rich literature proving undecidability of new problems within mathematics and computer science.  Undecidability in physics has also a natural continuation in the field of philosophy. However, in this review we only focus on mathematically rigorous results about well established models of physical systems, without entering into a discussion of their possible philosophical implications.

In this final outlook chapter we first provide a general view relating undecidability in mathematics and physics, discuss the scope of application of the results that we have covered, the consequences in physics that such an abstract concept as undecidability can have at a practical level in research, and some new directions that look promising for future research.

%%%%%%%%%%%%%%%%%%%%%%%%%%%%%%%%%%%%%%%%%%%%%%%%%%%%%%%%%%%%%%%%%%%%%%%%
%%%%%%%%%%%%%%%%%%%%%%%%%%%%%%%%%%%%%%%%%%%%%%%%%%%%%%%%%%%%%%%%%%%%%%%%
\subsection{General view}

\begin{figure}[h]
\noindent\makebox[\textwidth]{%
\includegraphics[width=15.5cm]{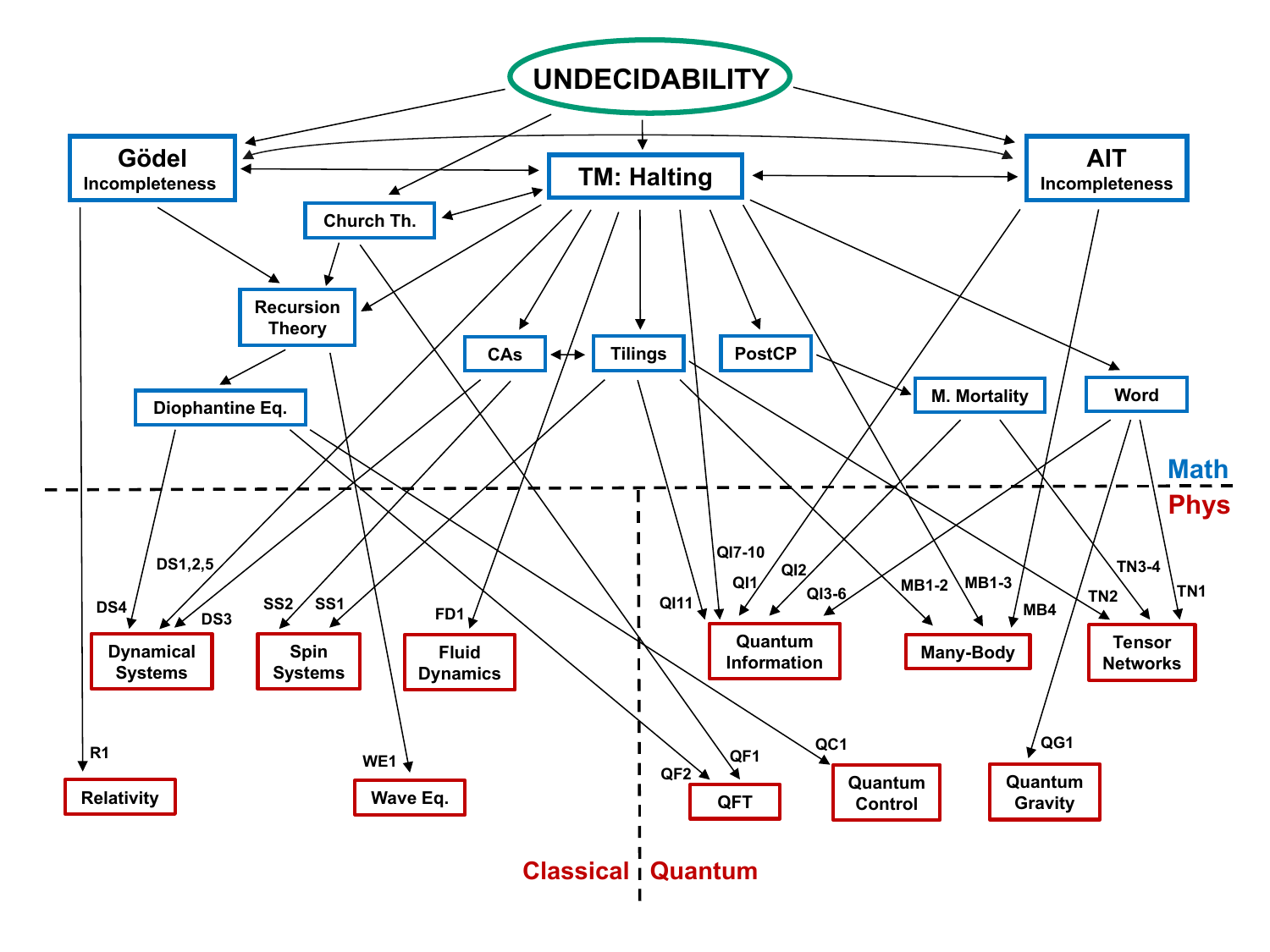}}
\centering
\caption{Diagram showing the connections between systems with undecidable properties in mathematics (blue) and physics (red - classical and quantum) that are commented on in this review (see main text).}
\label{fig:general}
\end{figure}

In Fig.\ \ref{fig:general} we provide a relational view of the results we have reviewed in this article. In the upper part (blue) we can see the undecidability results in mathematics that have been used in physics, stemming from the core ones: G\"odel, Turing, AIT. At the bottom (red) one can find the undecidability results in physics grouped into fields, with classical (quantum) systems on the left (right). Arrows come to each box from the mathematical systems that have been used to \textit{bring} undecidability from mathematics to physics. The labels on the arrows (e.g. DS4) indicate the specific article(s) that can be found in the following list.\\

\newpage

\textbf{CLASSICAL SYSTEMS}
\begin{itemize}

\item \textbf{Dynamical Systems}, section \ref{UDPFirst}:\\
-\textbf{DS1}. Fredkin and Toffoli (1982) \cite{Fredkin82}. \\
-\textbf{DS2}. Moore (1990) \cite{Moore90}. \\
-\textbf{DS3}. Omohundro (1984) \cite{Omohundro84}. \\
-\textbf{DS4}. da Costa and Doria (1991) \cite{daCosta91a, daCosta91b}. \\
-\textbf{DS5}. Blondel (2001) \cite{Blondel2001a, Blondel2001b}

\item \textbf{Spin Systems}: \\
-\textbf{SS1}. Kanter (1990) \cite{Kanter90}, section \ref{UDPFirst}.\\
-\textbf{SS2}. Gu et al.\ (2009); Gu and Perales (2012) \cite{Gu09, Gu12}, section \ref{MBClassical}.

\item \textbf{Fluid Dynamics}, section \ref{Fluid}:\\
-\textbf{FD1}. Cardona et al.\ (2021); Cardona et al.\ (2023)  \cite{Cardona20, cardona2023computability}.

\item \textbf{Relativity}, section \ref{UDPFirst}:\\
-\textbf{R1}. da Costa et al.\ (1990) \cite{daCosta1990}.

\item \textbf{Wave Equation}, section \ref{UDPFirst}:\\
-\textbf{WE1}. Pour-El and Richards (1981) \cite{Pourel81}. \\

\end{itemize}

\textbf{QUANTUM SYSTEMS}
\begin{itemize}

\item \textbf{Quantum Information}, section \ref{MBQI}: \\
-\textbf{QI1}.  Brukner (2009); Paterek et al.\ (2010); Trejo et al.\ (2023) \cite{Brukner09,Paterek10,Trejo2023}.\\
-\textbf{QI2}. Eisert et al.\ (2012) \cite{Eisert12}.\\
-\textbf{QI3}. Slofstra (2019, 2020) \cite{Slofstra19, Slofstra20}.\\
-\textbf{QI4}. Fritz (2021) \cite{Fritz16}.\\
-\textbf{QI5}. Mancinska and Roberson (2020) \cite{Mancinska20}.\\
-\textbf{QI6}. Fu et al.\ (2021) \cite{Fu21}.\\
-\textbf{QI7}. Elkouss and P\'erez-Garc\'ia (2018 \cite{Elkouss16}.\\
-\textbf{QI8}. Ji et al.\ (2020) \cite{Ji20}.\\
-\textbf{QI9}. Wright and Farkas (2023) \cite{wright2023invertible}.\\
-\textbf{QI10}. Bendersky et al.\ (2017) \cite{Bendersky16}.\\
-\textbf{QI11}. Van den Nest and Briegel (2008) \cite{VandenNest08}.

\item \textbf{Many-Body Systems}, section \ref{MBQuantum}:\\
-\textbf{MB1}. Cubitt et al.\ (2015, 2022); Bausch et al.\ (2021, 2020) \cite{Cubitt15c, Cubitt15b, Bausch19, Bausch20}.\\
-\textbf{MB2}. Watson et al.\ (2022) \cite{Watson21}.\\
-\textbf{MB3}. Shiraishi and Matsumoto (2021, 2020) \cite{Shiraishi20a, Shiraishi20b}.\\
-\textbf{MB4}. Purcell et al.\ (2024) \cite{Purcell24}.

\item \textbf{Tensor Networks}, section \ref{MBTN}: \\
-\textbf{TN1}. Morton and Biamonte (2012) \cite{Morton12}.\\
-\textbf{TN2}. Scarpa et al.\ (2020 \cite{Scarpa20}.\\
-\textbf{TN3}. Kliesch et al.\ (2014) \cite{Kliesch14}.\\
-\textbf{TN4}. De las Cuevas et al.\ (2016) \cite{Delascuevas16}.

\item \textbf{Quantum Field Theory}: \\
-\textbf{QF1}. Komar (1964) \cite{Komar64}, section \ref{UDPFirst}. \\
-\textbf{QF2}. Tachikawa (2023) \cite{Tachikawa23}, section \ref{QFT}.

\item \textbf{Quantum Control}, section \ref{Measurement}: \\
-\textbf{QC1}. Bondar and Pechen (2020) \cite{Bondar2020}.

\item \textbf{Quantum Gravity}, section \ref{UDPFirst}: \\
-\textbf{QG1}. Geroch and Hartle (1986) \cite{Geroch1986}.

\end{itemize}
\

Let us briefly discuss some conclusions that we can draw from Fig.\ \ref{fig:general}. First, we can appreciate the great diversity of physical systems in which undecidable properties have been proven, both quantum and classical, with most of these results obtained in the last 15 years. We also note that most of them introduce undecidability through mathematical systems derived directly or indirectly from the halting problem for Turing Machines, as in \textbf{DS1,2,5}, \textbf{FD1}, \textbf{QI7-10} and \textbf{MB1-3}. As mentioned above, one reason is that Turing Machines already resemble real physical systems, whereas G\"odel's or AIT arguments are of more abstract nature.

Using the halting problem as starting point, undecidability can be transferred to other mathematical systems by embedding a Turing Machine into them. In this way, one obtains undecidable properties in 2D mathematical systems like cellular automata and tilings that geometrically resemble 2D physical systems (classical or quantum spin systems, tensor networks), so it is possible to \textit{export}, directly or indirectly, undecidable problems, as in \textbf{DS3}, \textbf{SS1}, \textbf{SS2}, \textbf{MB1-2}, \textbf{TN2}, and \textbf{Q11}.

In addition to geometric similarity, the frontier between undecidability in mathematics and physics is crossed by means of other mathematical objects that appear naturally in physics, such as equations, matrices or groups. This is the case for \textbf{DS4}, \textbf{QI2}, \textbf{QI3-6}, \textbf{TN1,3,4}, \textbf{QC1}, \textbf{QG1}, \textbf{QF2}, which are based on Diophantine Equations, Matrix Mortality or the Word Problem.

There are a few results relying on less exploited connections in Fig.~\ref{fig:general}, such as \textbf{R1}, \textbf{WE1}, \textbf{QF1}, \textbf{QI1} and \textbf{MB4} . Indeed, the variety and richness of undecidable properties in mathematical systems (many of which are not in the figure) suggests that there may still be many more connections to be discovered that might prove undecidable properties in multiple and diverse physical systems.

%%%%%%%%%%%%%%%%%%%%%%%%%%%%%%%%%%%%%%%%%%%%%%%%%%%%%%%%%%%%%%%%%%%%%%%%
%%%%%%%%%%%%%%%%%%%%%%%%%%%%%%%%%%%%%%%%%%%%%%%%%%%%%%%%%%%%%%%%%%%%%%%%
\subsection{Infinite undecidable mathematics vs finite decidable nature}
\label{INF}

When studying a real physical system, by constraints in the measurement precision or in the sample size, one usually needs to consider only a finite number of possible configurations. Mathematical models to treat those situations as such are therefore decidable, since it suffices to consider one by one all the possible configurations. Take, for instance, 10 particles that are described by an effectively finite configuration space and that interact a finite number of times. Clearly, there is no way to embed a universal Turing Machine with its infinite tape into such a system.

What if instead of 10 particles we consider $10^{23}$?  It is still a decidable system, although practically intractable. When we take the number of particles to infinity, statistical mechanics and thermodynamics often allow us to model it in a simple and tractable way, and to make precise predictions about their collective behavior. However, as we have seen, there are cases where this limit is the midwife of undecidability.

In modeling nature we take limits in  other directions as well, which may be displayed by infinitesimal calculus, or asymptotic times---and again, these can either simplify the mathematical model or lead to undecidability.

From this point of view, undecidability seems to be less a feature of the physical system and more a property of the idealized mathematical model we use to describe it.

However, undecidability in the infinite limit often also has implications for the more realistic situation of a finite-size system. There are ``reflections'' of undecidability at finite system sizes. In particular, if undecidability is proven using an embedding of a universal Turing Machine, two potential consequences are lurking in the construction:
\begin{enumerate}
   \item The associated problem for finite size or finite precision becomes complete for some computational complexity class. Which complexity class will depend on the problem. For instance, the finite-size version of the problem to decide whether a Matrix Product Operator is semidefinite positive was proven to be NP-hard by \textbf{Kliesch et al.\ (2014)} \cite{Kliesch14}, while the problem of giving an estimate for the ground state energy density of a fixed quantum system in the thermodynamic limit, as a function of the number of bits of precision, was proven to be  hard for the complexity class ${\rm FEXP}^{\rm NEXP}$ (\textbf{Aharonov and Irani (2022); Watson and Cubitt (2022)} \cite{aharonov2022hamiltonian, watson2022computational}). It would be desirable to have general results that allow to infer finite size/precision complexity results based on the undecidability of its infinite counterpart. A first step in that direction was made by \textbf{Klingler et al.\ (2023)} \textbf{\cite{Klingler23}} and discussed further by \textbf{Wolf (2023)} \cite{Wolf23}.

    \item As we will discuss in more detail in section \ref{positiveside}, undecidability of infinite problems may imply unexpected and exotic size-dependent effects for their finite-size counterparts. That is, there may be a finite but uncomputably large size at which the properties of the system change abruptly and dramatically.
\end{enumerate}

%%%%%%%%%%%%%%%%%%%%%%%%%%%%%%%%%%%%%%%%%%%%%%%%%%%%%%%%%%%%%%%%%%%%%%
%%%%%%%%%%%%%%%%%%%%%%%%%%%%%%%%%%%%%%%%%%%%%%%%%%%%%%%%%%%%%%%%%%%%%%%%

\subsection{The effect of noise}
\label{noise}

There is yet another important ingredient to be considered, which is present in real physical systems: noise. For instance, it has been shown by \textbf{Dong et al.\ (2023)} \cite{dong2023computational}, culminating a series of previous works by \textbf{Qin and Yao (2021, 2023)} \cite{qin2021nonlocal, qin2023decidability}, that, under a particular type of noise, where the players are allowed to share arbitrarily many copies of a single state with a fixed amount of noise per state, the undecidability of approximating the quantum value of a non-local game downgrades to (containment in) the complexity class NEXP. It is an important open question to understand which undecidability results are stable under natural sources of noise in the corresponding physical problem; and, in case they are not stable, to determine the exact complexity class to which their noisy versions belong.

%%%%%%%%%%%%%%%%%%%%%%%%%%%%%%%%%%%%%%%%%%%%%%%%%%%%%%%%%%%%%%%%%%%%%%
%%%%%%%%%%%%%%%%%%%%%%%%%%%%%%%%%%%%%%%%%%%%%%%%%%%%%%%%%%%%%%%%%%%%%%%%
\subsection{Undecidability over the real numbers}

So far, as becomes clear from this review, the focus of the research in the area of undecidability in physics has been to prove that concrete relevant problems in the physics literature are undecidable / uncomputable in the Turing sense.

However, in the seminal book by \textbf{Blum et al.\ (1998)} \cite{Blum98}, motivated by optimization algorithms like Newton's algorithm, the paradigm of computation over the real numbers (instead of the integers) is formalized. In this paradigm, decidable sets are tightly connected with algebraic sets (i.e.\ zeroes of polynomials). In particular, this allows to show that membership in fractal sets like the Mandelbrot set are undecidable in this different computational paradigm.

The existence of fractal sets in many interesting physical systems, such as Hofstadter's Butterfly \cite{hofstadter1976energy}, makes it tempting to suggest that undecidability over the reals could also be of relevance in physics.

%%%%%%%%%%%%%%%%%%%%%%%%%%%%%%%%%%%%%%%%%%%%%%%%%%%%%%%%%%%%%%%%%%%%%%%%
%%%%%%%%%%%%%%%%%%%%%%%%%%%%%%%%%%%%%%%%%%%%%%%%%%%%%%%%%%%%%%%%%%%%%%%%

\subsection{The positive side}
\label{positiveside}

Undecidability puts limits on our ability to reason mathematically about a system. But it also has a positive side. It shows the existence of ``pathological'' examples that can provide counterexamples to open problems in physics. That is, undecidability can be used as a proof technique. We have already seen a couple of examples in this spirit. In section \ref{MBTN}, this technique is used to show that there exist Matrix Product Density Operators without a local purification, a question that goes back to \cite{Fannes92}. In a similar vein, undecidability of the ground state energy density has been used as a proof element by \textbf{Blakaj and Wolf (2024)} \cite{BlakajWolf23} in a discussion of the set of reduced states of infinite quantum spin chains. But it is in section \ref{QNL} where this technique shows its real power, providing the solution to Tsirelson's problem and, even more remarkably, to Connes' embedding problem.

Moreover, those pathological examples could be interesting on their own, since they can display new physics. One illustrative example in this direction is the existence of {\it size-driven quantum phase transitions} by \textbf{Bausch et al.\ (2018)} \cite{Bausch2018}, a new type of quantum effect that is a consequence of the undecidability of the spectral gap: systems whose thermodynamic properties in the infinite limit differ radically from the properties they display at any finite size below a threshold that can be in principle as large as desired. Finding analogous size-driven results for other undecidable problems could be very interesting. This could include for instance a communication channel whose properties change abruptly after a fixed (but arbitrary large) number of uses. (A step in this direction is shown in \cite{Cubitt15c}, but there the result concerns a family of channels of increasing dimension rather than a single, fixed communication channel.) Or a non-local game for which the quantum value enjoys a fixed jump if played with quantum entanglement dimension larger than a fixed (but arbitrary large, possibly uncomputable) number.

\section*{Acknowledgments}

DPG acknowledges support from the Spanish Ministry of Science and Innovation MCIN/AEI/10.13039/501100011033 (grants CEX2023-001347-S and PID2020-113523GB-I00) and from Universidad Complutense de Madrid (grant FEI-EU-22-06). MG acknowldges support from the CQT Bridging Grant, the Singapore Ministry of Education Tier 1 Grants RG77/22 and RT4/23, the Singapore Ministry of Education Tier 2 Grant MOE-T2EP50221-0005, grant no.~FQXi-RFP-1809 (The Role of Quantum Effects in Simplifying Quantum Agents) from the Foundational Questions Institute and Fetzer Franklin Fund (a donor-advised fund of Silicon Valley Community Foundation). This research was supported in part by Perimeter Institute for Theoretical Physics. Research at Perimeter Institute is supported by the Government of Canada through the Department of Innovation, Science, and Economic Development, and by the Province of Ontario through the Ministry of Colleges and Universities. MMW has been supported by the Deutsche Forschungsgemeinschaft (DFG, German Research Foundation) under Germany's Excellence Strategy EXC-2111 390814868, and the SFB/Transregio TRR 352 – Project-ID 470903074.

\bibliography{UDP-PR-21.10.24}

\end{document}